\documentclass[prd,nofootinbib,preprint,superscriptaddress,twocolumn,10pt]{revtex4}
\pdfoutput=1
\usepackage{aas_macros}
\usepackage{newtxtext,newtxmath}
\usepackage{hyperref}
\usepackage{booktabs}
\usepackage{mathtools}
\usepackage{subfigure}
\usepackage{bm}
\usepackage{ulem}
\usepackage{orcidlink}
\newcommand{\be}{\begin{equation}}
\newcommand{\ee}{\end{equation}}
\newcommand{\ba}{\begin{eqnarray}}
\newcommand{\ea}{\end{eqnarray}}
\newcommand{\bi}{\begin{itemize}}
\newcommand{\ei}{\end{itemize}}

\newcommand{\bmcinv}{[\mathbf{N}^{-1}]}
\usepackage{color} 

\newcommand{\red}{\textcolor{red} }

\newcommand{\reffig}{Fig.\ref}

\usepackage{xcolor}
\DeclareRobustCommand{\VAN}[3]{#2}
\let\VANthebibliography\thebibliography
\def\thebibliography{\DeclareRobustCommand{\VAN}[3]{##3}\VANthebibliography}
\usepackage{graphicx}
\usepackage{grffile}

\begin{document}
\title{{An Attempt to Reconstruct Weak Lensing by Counting DECaLS Galaxies}}

\author{Jian Qin\orcidlink{0000-0003-0406-539X}}
\email{qinjian@sjtu.edu.cn}
\affiliation{Department of Astronomy, School of Physics and Astronomy, Shanghai Jiao Tong University, Shanghai, 200240,China}
\affiliation{Key Laboratory for Particle Astrophysics and Cosmology
(MOE)/Shanghai Key Laboratory for Particle Physics and Cosmology,China}

\author{Pengjie Zhang\orcidlink{0000-0003-2632-9915}}
\email{zhangpj@sjtu.edu.cn}
\affiliation{Department of Astronomy, School of Physics and Astronomy, Shanghai Jiao Tong University, Shanghai, 200240,China}
\affiliation{Tsung-Dao Lee Institute, Shanghai Jiao Tong University, Shanghai, 200240, China}
\affiliation{Key Laboratory for Particle Astrophysics and Cosmology
(MOE)/Shanghai Key Laboratory for Particle Physics and Cosmology,China}

\author{Haojie Xu}
\affiliation{Shanghai Astronomical Observatory, Chinese Academy of Sciences, Shanghai 200030, China}
\affiliation{Department of Astronomy, School of Physics and Astronomy, Shanghai Jiao Tong University, Shanghai, 200240,China}
\affiliation{Key Laboratory for Particle Astrophysics and Cosmology
(MOE)/Shanghai Key Laboratory for Particle Physics and Cosmology,China}

\author{Yu Yu\orcidlink{0000-0002-9359-7170}}
\email{yuyu22@sjtu.edu.cn}
\affiliation{Department of Astronomy, School of Physics and Astronomy, Shanghai Jiao Tong University, Shanghai, 200240,China}
\affiliation{Key Laboratory for Particle Astrophysics and Cosmology
(MOE)/Shanghai Key Laboratory for Particle Physics and Cosmology,China}

\author{Ji Yao}
\affiliation{Shanghai Astronomical Observatory, Chinese Academy of Sciences, Shanghai 200030, China}
\affiliation{Department of Astronomy, School of Physics and Astronomy, Shanghai Jiao Tong University, Shanghai, 200240,China}
\affiliation{Key Laboratory for Particle Astrophysics and Cosmology
(MOE)/Shanghai Key Laboratory for Particle Physics and Cosmology,China}

\author{Ruijie Ma}
\affiliation{Department of Astronomy, School of Physics and Astronomy, Shanghai Jiao Tong University, Shanghai, 200240,China}
\affiliation{Key Laboratory for Particle Astrophysics and Cosmology
(MOE)/Shanghai Key Laboratory for Particle Physics and Cosmology,China}

\author{Huanyuan Shan}
\affiliation{Shanghai Astronomical Observatory, Chinese Academy of Sciences, Shanghai 200030, China}

\begin{abstract}
{In this study, we explore the reconstruction of the lensing convergence ${\kappa}$ map using the cosmic magnification effect in DECaLS galaxies from the DESI imaging surveys DR9. This complementary approach provides valuable insights into the large scale structure of the Universe alongside traditional cosmic shear measurements.} It is achieved by linearly weighing $12$ maps of galaxy number overdensity in different magnitude bins of $grz$ photometry bands. The weight is designed to eliminate the mean galaxy deterministic bias, minimize galaxy shot noise while maintaining the lensing convergence signal. We also perform corrections of imaging systematics in the galaxy number overdensity. The $\hat{\kappa}$ map has $8365$ deg$^2$ sky coverage. Given the low number density of DECaLS galaxies, the $\hat{\kappa}$ map is overwhelmed by shot noise and the map quality is difficult to evaluate using the lensing auto-correlation. {Alternatively, we measure its cross-correlation with the cosmic shear catalogs of DECaLS galaxies of DESI imaging surveys DR8, which have a sky area that fully covers the reconstructed $\hat{\kappa}$ map.}  {Since the $\hat{\kappa}$ map is contamined by residual galaxy clustering, we model it as $\hat{\kappa}=A\kappa+\epsilon\delta_{\rm m}+\cdots$.
Under this assumption, we detect a convergence-shear cross-correlation signal with $S/N\simeq 22$. } The analysis also shows that  the  galaxy intrinsic clustering is suppressed by a factor $\mathcal{O}(10)$. Various tests with different galaxy and shear samples, and the Akaike information criterion analysis all support the lensing detection.  So is the  imaging systematics corrections, which enhance the lensing signal detection by $\sim 30\%$.   We discuss various issues for further improvement of the measurements. 
\end{abstract}
\maketitle

\section{Introduction}

Weak lensing, which probe directly the matter distribution of the Universe, provide powerful insight into dark energy, dark matter and gravity at cosmological scales \citep{2001PhR...340..291B, 2015RPPh...78h6901K, HJ2008, van2010a, FuFan2014}. 
An effect of weak lensing is cosmic shear, which distorts shapes of distant galaxies. This signal can be
extracted statistically from the galaxy images and has been the main target of current weak lensing studies. 
With large surveys, such as the on-going Dark Energy Survey \citep[DES,][]{DES2016}, the Kilo-Degree Survey \citep[KiDS,][]{KiDS2013}, 
and the Hyper Suprime-Cam Subaru Strategic Program survey\citep[HSC-SSP,][]{HSC2018}, Vera C. Rubin Observatory's Legacy Survey of Space and Time \citep[LSST,][]{2009arXiv0912.0201L} and Euclid \citep{2011arXiv1110.3193L}, cosmic shear contributed significantly to the precision cosmology  
\citep[e.g.,][]{Hamana2020,2021A&A...645A.104A,2021A&A...645A.105G, 2022A&A...665A..56L,PhysRevD.105.023514,PhysRevD.105.023515, 2023arXiv230400702L}.

Another weak lensing effect is cosmic magnification, which induces changes in the observed galaxy number density by magnifying the galaxy flux and resizing the solid angle of the sky patches \citep{BN1992, BN1995}.
As the galaxy intrinsic alignment contaminates the cosmic shear signal, the galaxy intrinsic clustering contaminates the cosmic magnification signal. 
In observations, cosmic magnification is typically measured through cross correlations of two samples in the same patch of the sky but widely separated in redshift. 
{Lenses, located at lower redshifts, include luminous red galaxies (LRGs) and clusters \citep[e.g.,][]{pub.1059915677,2019MNRAS.484.1598B,2016MNRAS.457.3050C,2020MNRAS.495..428C}. Sources, located at higher redshifts, include quasars \citep{2005ApJ...633..589S,2012ApJ...749...56B}, Lyman break galaxies \citep{2012MNRAS.426.2489M,2017A&A...608A.141T},  and submillimetre galaxies \citep{2021A&A...656A..99B,Crespo_GonzalezNuevo_Bonavera_Cueli_Casas_Goitia_2022}.}
In addition to this, the magnification effects can be detected through the shift in number count, magnitude and size \citep{2010MNRAS.405.1025M,2011MNRAS.411.2113J,Schmidt_Leauthaud_Massey_Rhodes_George_Koekemoer_Finoguenov_Tanaka_2012,2014ApJ...780L..16H,Duncan_Heymans_Heavens_Joachimi_2016,2018MNRAS.476.1071G}. Recently, cross-correlation between cosmic magnification and cosmic shear has been detected in HSC \citep{Liu_Liu_Gao_Wei_Li_Fu_Futamase_Fan_2021} and DECaLS $\times$ DESI \citep{2023MNRAS.524.6071Y}.

 {These are the indirect measurements of cosmic magnification through cross-correlation.
 It is in principle possible to extract the magnification signal directly from multiple galaxy overdensity maps of different brightness} \citep{2005PhRvL..95x1302Z,2011MNRAS.415.3485Y,ABS,YangXJ15,YangXJ17,Zhang18,2021RAA....21..247H}. The key information to use here is the characteristic flux dependence of magnification bias. The major contamination to eliminate is the intrinsic galaxy clustering. The galaxy bias is complicated \citep[e.g.][]{Bonoli09,Hamaus10,Baldauf10}, however the leading order component to eliminate is the deterministic bias.  \cite{2021RAA....21..247H} proposed a modified internal linear combination (ILC) method, which can eliminate the average galaxy bias model-independently. 

We extend the methodology of  \cite{2021RAA....21..247H} to utilize multiple photometry band information. Including the multi-band information not only improves the mitigation of galaxy intrinsic clustering, but also suppresses shot noise, as recently shown by \cite{2024MNRAS.527.7547M}.  We then apply it to DECaLS galaxies of the DESI imaging surveys DR9. The paper is organized as follows. In Sec. \ref{sec:method2}, we present the reconstruction method and the modeling of the cross correlation. In Sec. \ref{sec:data3}, we describe the data and how we process the data, which include the galaxy samples for lensing reconstruction, the imaging systematics mitigation, the galaxy samples with shear measurement and the cross correlation measurement.  Sec. \ref{sec:result4} contains details of the cross-correlation analyses including the fitting to the model and the internal tests of the analysis.
Summary and discussions are given in Sec. \ref{sec:summary5}.

\section{Method}\label{sec:method2}
\subsection{Lensing convergence map reconstruction }
We aim to conduct lensing reconstruction for DECaLS galaxies based on the ideas proposed in \cite{2005PhRvL..95x1302Z,2011MNRAS.415.3485Y,2021RAA....21..247H}.  In addition to galaxy flux, we leverage the information provided by the galaxy photometry bands.  Specifically, we utilize the data obtained from the $g$, $r$, and $z$ bands and sort the galaxies into $N_F$ flux bins for each band.
In the weak lensing regime, the galaxy number overdensity of each flux bin has \citep[e.g.][]{scranton2005detection,2011MNRAS.415.3485Y}
\begin{equation} \label{delta}
\delta^L_{i}=
b_i\delta_m+g_i\kappa+\delta_i^S\ .
\end{equation}
Here $\delta_m$ is the underlying dark matter overdensity. $b_i$ and $g_i$ are the deterministic bias and the magnification coefficient in the $i$-th flux bin. $\delta_i^S$ denotes the term from galaxy stochasticity. 
Cosmic magnification modulates the galaxy density field by $g\kappa$, where $\kappa$ is the lensing convergence.
{In other words, $g\kappa$ represents the response of the galaxy number overdensity $\delta_i^L$ to weak lensing, with the linear factor $g$ commonly referred to as the magnification coefficient.
The value of $g$ depends on the galaxy selection criteria and observational conditions \citep{Wietersheim-Kramsta_Joachimi_van, JElvinPoole2022DarkES}.
For a flux-limited sample / selection criteria, where the only selection is that galaxies brighter than a certain flux threshold are included, 
$g$ is determined by the logarithmic slope of the galaxy luminosity function} \citep[e.g.][]{10.1093/mnras/stad1594,2024MNRAS.527.1760W}
\begin{equation}
    g=2(\alpha - 1)\ ,\ \ \ \alpha=-\frac{d\ln{n(F)}}{d\ln{F}}-1\ .
    \label{eq:prefactorg}
\end{equation}
For a source at redshift $z_s$, $\kappa$ directly probes the underlying matter overdensity $\delta_m$ by \citep[e.g.][]{Liu_Liu_Gao_Wei_Li_Fu_Futamase_Fan_2021,2023MNRAS.524.6071Y}
\begin{equation}
\kappa\left(\boldsymbol{\theta}, z_{\mathrm{s}}\right)=\frac{3 H_{0}^{2} \Omega_{\mathrm{m}}}{2 c^{2}} \int_{0}^{\chi_{\mathrm{s}}} \frac{D\left(\chi_{\mathrm{s}}-\chi\right) D(\chi)}{D\left(\chi_{\mathrm{s}}\right)} \delta_{\mathrm{m}}(z, \boldsymbol{\theta})(1+z) d \chi\ .
\end{equation}
Where $\chi$ and $\chi_{s}$ are the radial comoving distances to the lens at redshift $z$ and the source at redshift $z_{\mathrm{s}}$, respectively. $D(\chi)$ denotes the comoving angular diameter distance, which equals $\chi$ for a flat universe.

A linear estimator of the convergence $\kappa$ has the form \citep{2021RAA....21..247H}
\be\label{eq:linear combination}
\hat{\kappa}=\sum_{i} w_{i}\delta_{i}^{\rm L}\ , 
\ee 
The weight $w_i$ is determined by minimizing the shot noise, 
\be \label{eqn:shot}
\left\langle\left|\sum_{i,j}
w_{i}w_{j}\delta_{ij}^{\rm shot}\right|^2\right\rangle=\sum_{i,j}
w_{i}w_{j}\frac{\bar{n}_{ij}}{\bar{n}_{i}\bar{n}_{j}}\ . \ee 
under conditions, \be \label{eqn:wg} \sum_{i} w_{i}g_{i}=1\ ,
\ee \be \label{eqn:wb} \sum_{i} w_{i}=0\ . \ee 
Here, $\bar{n}_{i}$ is the
average galaxy surface number density of the $i$-th flux bin, while $\bar{n}_{ij}$ is that both in the $i$-th and $j$-th flux bin. 
$\bar{n}_{ij}=0$ if $i$-th and the $j$-th flux bin are from the same photometry band.
Using the Lagrangian
multiplier method, the solution is
\begin{equation}
    w_{i}=\frac{1}{2} \left( \lambda_{1}\sum_{j=1}^{N_F}\bmcinv_{ij}g_{j}+\lambda_{2}\sum_{j=1}^{N_F}\bmcinv_{ij}  \right)~ ,
\label{eq:weights}
\end{equation}
where $\mathbf{N}$ is the matrix of $\frac{\bar{n}_{ij}}{\bar{n}_i\bar{n}_j}$ and the two Lagrangian multipliers are given by
\begin{subequations}
\begin{equation}
    \lambda_{1}=\frac{2\sum_{i,j=1}^{N_F} \bmcinv_{ij}}{(\sum_{i,j=1}^{N_F}\bmcinv_{ij}g_{i}g_{j})\cdot(\sum_{i,j=1}^{N_F} \bmcinv_{ij})-(\sum_{i,j=1}^{N_F} \bmcinv_{ij}g_{i})^{2}}~ , 
\end{equation}
\begin{equation}
    \lambda_{2}=\frac{-2\sum_{i,j=1}^{N_F} \bmcinv_{ij}g_{j}}{(\sum_{i,j=1}^{N_F}\bmcinv_{ij}g_{i}g_{j})\cdot(\sum_{i,j=1}^{N_F} \bmcinv_{ij})-(\sum_{i,j=1}^{N_F} \bmcinv_{ij}g_{i})^{2}}~ . 
\end{equation}
\label{eq:multipliers}
\end{subequations}
Plugging the above weight into Eq.\eqref{eq:linear combination}, we obtain the reconstructed/estimated lensing convergence $\hat{\kappa}$. 

\subsection{Further mitigation through the  convergence-shear cross-correlation}
The reconstructed map can then be expressed as 
\be\label{eq:k+em}
\hat{\kappa}=A\kappa+\epsilon\delta_{\rm m}+\cdots\ .
\ee
There are two important issues to address.  One is the residual galaxy clustering in the reconstructed $\hat{\kappa}$. The deterministic bias $b_i$, after weighting, becomes
\begin{equation}
    \epsilon=\sum_i w_i b_i\ .
\end{equation}
The other is a potential multiplicative error in the overall amplitude of $\hat{\kappa}$, which can arise from measurement error/bias in $g=2(\alpha - 1)$.  We quantify it with a dimensionless parameter $A$. 
{We aim to construct a flux-limited sample from the DECaLS galaxies to ensure that $g$, estimated by the logarithmic slope of the luminosity function (Eq. \ref{eq:prefactorg}), remains unbiased.
However, this is not always guaranteed, as observational conditions may introduce additional galaxy selection, such as the instrument or the redshift failure.
Fully accounting for these observational conditions would enable a more accurate estimation of $g$, which is beyond the scope of this work.
In this work, we use the $g$ estimated by Eq. \ref{eq:prefactorg} as an approximation.
}
The neglected terms in Eq. \ref{eq:k+em} include stochastic galaxy bias, shot noise, etc. 
An ideal estimator would achieve $A=1$ and $\epsilon=0$. But our estimator only guarantees $\sum_i w_i=0$ (Eq. \ref{eqn:wb}), so we have to explicitly check whether $\epsilon\neq 0$.\footnote{This issue can be solved by the principal component analysis of the galaxy cross-correlation matrix in hyperspace of galaxy properties \citep{Zhou_Zhang_Chen_2023,2023arXiv230615177M}. However this method requires robust clustering measurements, inapplicable to the DR9 galaxies that we use. } Also, given uncertainties in the estimation of $g=2(\alpha - 1)$ \citep{Wietersheim-Kramsta_Joachimi_van,JElvinPoole2022DarkES}, we must evaluate $A$ by the data and check whether $A=1$. 

Motivated by \cite{Liu_Liu_Gao_Wei_Li_Fu_Futamase_Fan_2021}, we propose to cross-correlate our $\hat{\kappa}$ map with cosmic shear catalogs in the same patch of sky, but of multiple redshift bins. In the cross-correlation, the neglected  in Eq. \ref{eq:k+em}, such as stochastic galaxy bias, do not contaminate\footnote{{The stochastic galaxy bias represents the portion of galaxy clustering that is not correlated with the cosmic density field (at the two point level). It does not correlate with cosmic shear, as cosmic shear is a direct result of the gravitational field generated by the inhomogeneities in the cosmic density field. Therefore, it does not contribute to the (two point) cross-correlation signal.}}. We then have the convergence-tangential shear correlation prediction, 
\be\label{eq:xikg_decom}
{\xi}^{\kappa\gamma}_{j,\rm th}(\theta)=
A\xi^{\kappa\gamma}_j(\theta)+\epsilon\xi^{m\gamma}_j(\theta)\ .
\ee
The above correlation is the convergence/matter-tangential shear correlation. $\theta$ is the angular separation and $j$ denotes the $j$-th source redshift bin of cosmic shear catalog.  Since $A$ and $\epsilon$ do not vary with $j$, we can simultaneously constrain cosmological parameters together with $A$ and $\epsilon$,  through measurements of $\hat{\xi}_j^{\kappa \gamma}$ at various source redshift bins. For the current work with the primary goal to test the feasibility of our reconstruction method, we fix the cosmology as the bestfit Planck 2018 flat $\Lambda$CDM cosmology \citep{Planck2018parameters} with key cosmological parameters $\Omega_m = 0.315, \Omega_\Lambda =1-\Omega_m, n_s = 0.965, h = 0.674$ and $\sigma_8 = 0.811$.  

We calculate the theoretical $\xi^{\kappa \gamma}$ and $\xi^{m\gamma}$ under the Limber approximation \citep{Limber1953}. 
The correlation functions are related to the corresponding power spectra via
\be\label{eq:kg:hankel}
\xi^{\kappa(m)\gamma}(\theta)=\int_0^\infty\frac{d\ell\ell}{2\pi}C_\ell^{\kappa(m)\gamma}J_2(\ell\theta)\ .
\ee
Here $J_2$ is the 2nd order Bessel function. $C_\ell^{\kappa\gamma}$ and $C_\ell^{m\gamma}$ are the shear-convergence and shear-matter cross power spectrum. 
In a flat Universe, they are expressed by
\be
    C_\ell^{\kappa(m)\gamma} =\int d\chi W^{\kappa(m)}W^\gamma(\chi)(\chi)\frac{1}{\chi^2}P_{m}\left(k = \frac{\ell}{\chi};z\right) \ .
\label{eqn:ckg}
\ee
Here $\chi$ is the comoving radial distance and $P_m$ is the 3-dimensional matter power spectrum. $W^{\kappa}$,$W^{\gamma}$ and $W^m$ are the projection kernels,
\be
    W^{\kappa(\gamma)}(\chi) = \frac{3H_0^2\Omega_{m}}{2a(\chi)c^2}
    \int d\chi n_{\rm len}(\chi)\frac{\chi-\chi^\prime}{\chi}\ .
\ee
\be
    W^m(z) = n_m(z) = \frac{c}{H(z)}W^m(\chi) \ ,
\ee
where $n_{\rm len}$ and $n_m$ is the normalized redshift distribution of the lens and dark matter tracers.

Then by fitting against ${\xi}_{j,\rm th}^{\kappa \gamma}$ at multiple shear redshifts, we constrain both $A$ and $\epsilon$. The two then quantify the convergence map quality from both the viewpoint of detection significance ($A/\sigma_A$), and systematic errors ($\epsilon$). 

\begin{figure} 
\centering
\includegraphics[width=0.9\columnwidth]{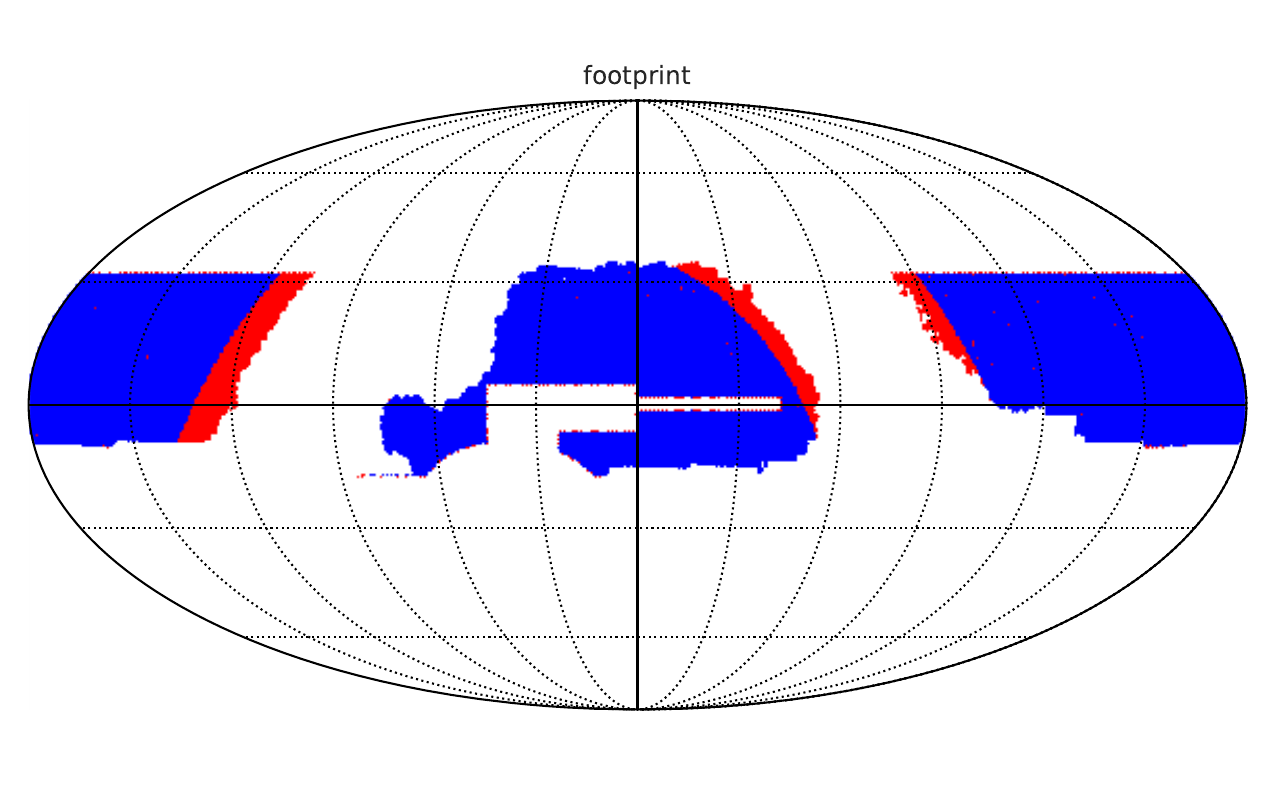}
\caption{The footprints of the DECaLS galaxies of LS DR9 used for weak lensing reconstruction (blue) and  the DECaLS galaxies of LS DR8 with shear measurement (red).
The two catalogs have a footprint overlap of $8365\deg^2$.
\label{fig:footprint}}
\end{figure}

\begin{table*}
    \centering
    \begin{tabular}{cccccccc}
    \hline
    \hline
{Subsample} &  Magnitude Range &     Galaxy Number      &  $\bar{n}({\rm arcmin}^{-2})$ &  $\alpha$ &     $g$ &     $w$ & band \\
\midrule
1  &    (23.4, 23.5) & 5959428.0 &         0.20 &   2.13 &  2.25 &  0.05 &    g \\
2  &    (23.2, 23.4) & 5959428.0 &         0.20 &   2.32 &  2.65 &  0.07 &    g \\
3  &    (22.9, 23.2) & 5959428.0 &         0.20 &   2.59 &  3.18 &  0.09 &    g \\
4  &    (\ , 22.9)   & 5959428.0 &         0.20 &   3.08 &  4.17 &  0.13 &    g \\
\hline
5  &    (22.9, 23.0) & 4217719.0 &         0.14 &   2.26 &  2.53 &  0.01 &    r \\
6  &    (22.7, 22.9) & 4217719.0 &         0.14 &   2.35 &  2.71 &  0.00 &    r \\
7  &    (22.4, 22.7) & 4217719.0 &         0.14 &   2.36 &  2.73 & -0.02 &    r \\
8  &    (\ , 22.4)   & 4217719.0 &         0.14 &   2.87 &  3.74 & -0.03 &    r \\
\hline
9  &    (22.3, 22.5) & 13088310.0 &         0.43 &   0.75 & -0.51 & -0.13 &    z \\
10 &   (22.0, 22.3)  & 13088310.0 &         0.43 &   1.10 &  0.20 & -0.10 &    z \\
11 &    (21.6, 22.0) & 13088310.0 &         0.43 &   1.46 &  0.91 & -0.08 &    z \\
12 &   (\ , 21.6)    & 13088310.0 &         0.43 &   2.18 &  2.35 & -0.01 &    z \\

    \hline
    \end{tabular}
    \caption{ Summary of the galaxy sub-samples used for lensing reconstruction. 
    We use galaxies in the photo-$z$ range $0.9<z_\kappa<1.2$.
    Throughout the paper we use $z_\kappa$ to denote the (photometric) redshift of galaxies 
    used for lensing reconstruction, 
    and use $z_\gamma$ for redshift of galaxies for cosmic shear measurements. 
    In addition to the information presented in the table, it should be noted that there are overlaps between the sub-samples in terms of the galaxies they contain.
    For example, sample 1 and sample 5 share a common set of $1.7\times10^6$ galaxies, which we denote as $n_{15}=3.4\times10^6$.
    Similarly $n_{16}=2.7\times10^6$ and $n_{17}=1.4\times10^6$.
    However $n_{ij}=0$ if sample i and j are from the same photometry band (e.g. $n_{12}=n_{13}=0$).
    }
    \label{tab:sub-sample}
\end{table*}


\begin{figure*} 

\centering
\subfigure{\includegraphics[width=0.67\columnwidth]{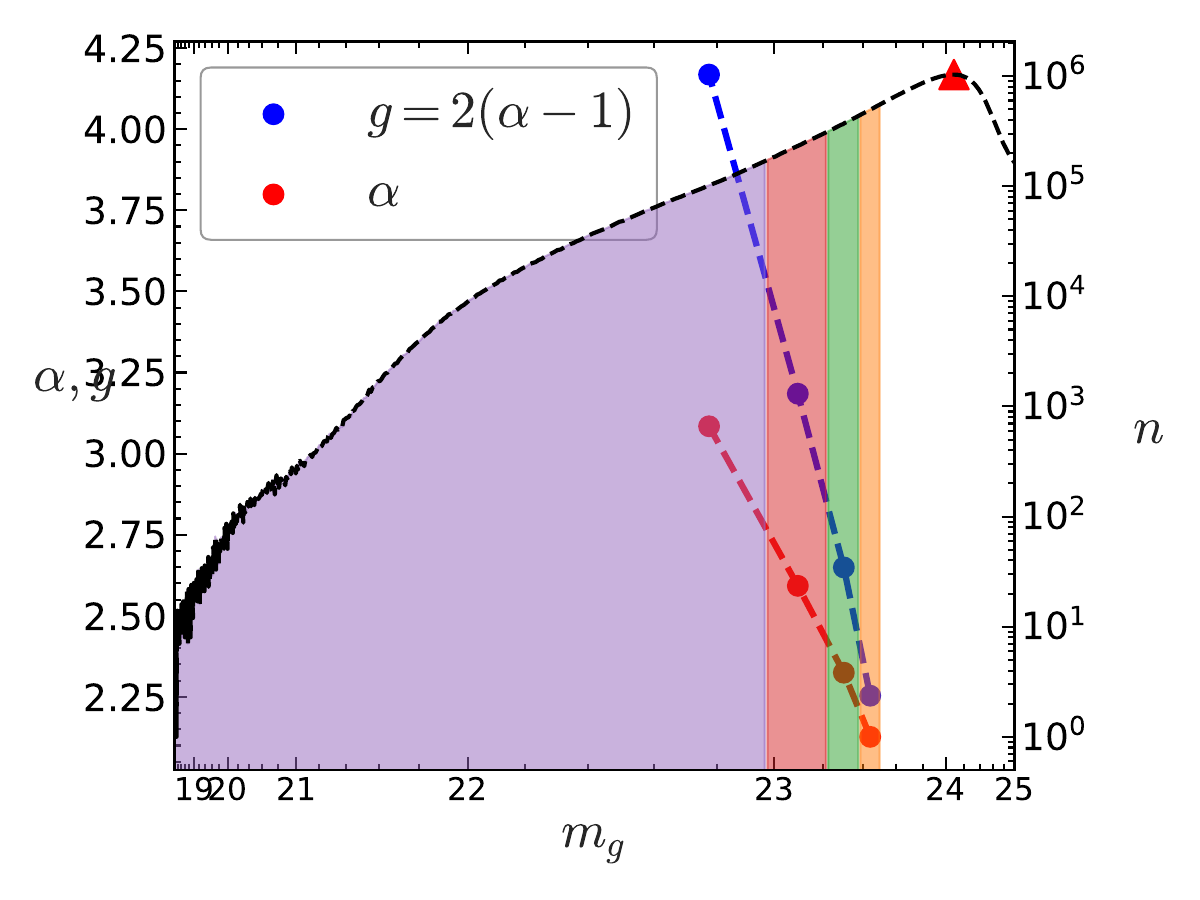}}
\centering
\subfigure{\includegraphics[width=0.67\columnwidth]{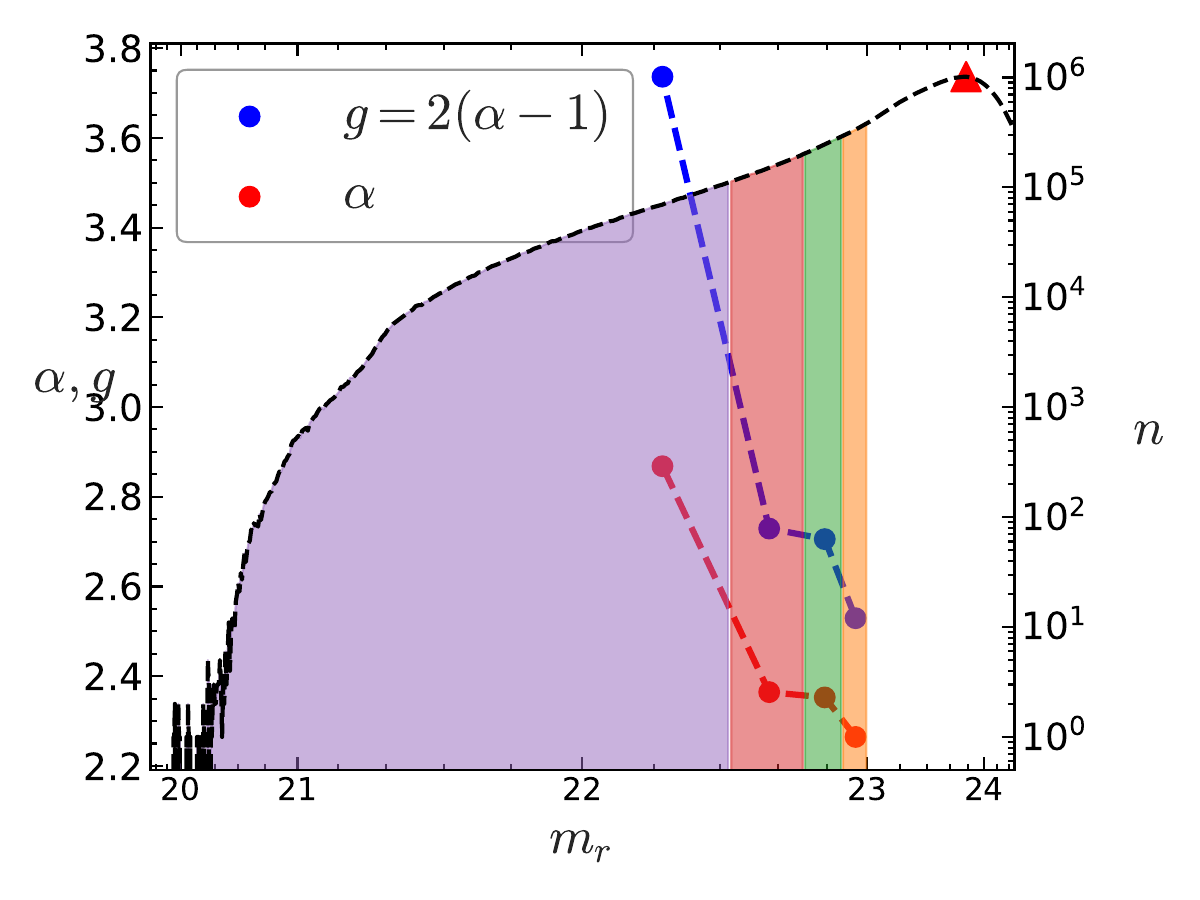}}
\centering
\subfigure{\includegraphics[width=0.67\columnwidth]{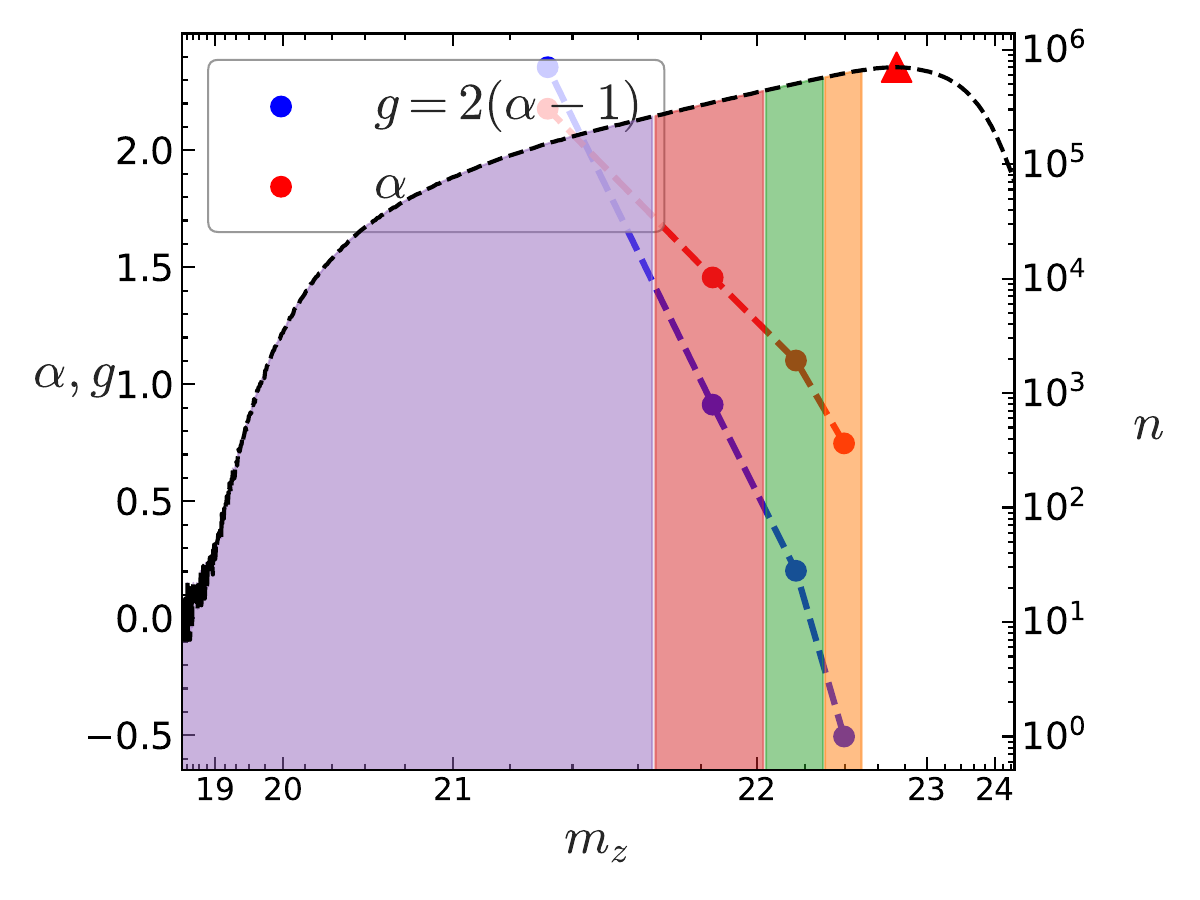}}
\caption{The galaxy number counts as a function of magnitude in the three bands ($g$, $r$, and $z$ from left to right panels), shown by the black dashed lines. 
To reconstruct the lensing convergence map, we divide the galaxies of each band into four flux bins, indicated by the four colored regions.
To ensure a reliable flux limit selection, the largest magnitude cuts are set to be $\sim0.5$ magnitude lower than the peak position (denoted by the red triangular points) of the galaxy number counts.
Within each flux bin, $\alpha$ and $g$ factors are calculated by Eq.\eqref{eq:a=dlogn_dlogf}.
And the results of $\alpha$ and $g$ are shown by the red and blue round points, respectively.
The left y-axis denote the values of $\alpha$ and $g$ factors, while the right y-axis denote the galaxy number count.
The results of $\alpha$ and $g$ are different in different bands, this source of cosmic magnification information help to improve the lensing reconstruction.
Note that the x-axis is not uniformly scaled, in order for a clearer visualization.
\label{fig:a,g}}
\end{figure*}

\section{Data analysis}\label{sec:data3}

We apply our method of lensing reconstruction to the DECaLS galaxies of the DESI imaging surveys \citep{dey2019overview} Data Release 9 (LS DR9). 
\subsection{Data}
The DECaLS data are processed by Tractor \citep{lang2016wise,meisner2017deep}. 
The galaxy samples used for lensing reconstruction are created in accordance with the selection criteria outlined in section 2.1 of \cite{yang2021extended}.
We summarize the main steps here. First, we select out extended imaging objects according to the morphological types provided by the TRACTOR software \citep{lang2016tractor}. 
We choose objects that have been observed at least once in each optical band to ensure a reliable photo-$z$ estimation.
We also remove the objects within $|\mathrm{b}|<25.0^{\circ}$ (where $\mathrm{b}$ is the Galactic latitude) to avoid high stellar density regions. Finally, we remove any objects whose fluxes are affected by the bright stars, large galaxies, or globular clusters (maskbits 1,5,6,7,8,9,11,12,13\footnote{https://www.legacysurvey.org/dr9/bitmasks/} ). The sky coverage of our selected DECaLS sample is shown in \reffig{fig:footprint}. We apply identical selections to the publicly available random catalogues \footnote{https://www.legacysurvey.org/dr9/files/\#random-catalogues-randoms}.

For the convergence-shear cross-correlation analysis, we utilize the shear measurements from DECaLS galaxies of DESI imaging surveys DR8 (LS DR8), with a sky coverage of $8960\deg^2$.
The footprint of the DECaLS galaxies of LS DR8 is shown in \reffig{fig:footprint}. 
These galaxies are then divided into five types according to their morphologies: PSF, SIMP, DEV, EXP, and COMP.
The ellipticity $e_{1,2}$ are estimated -- except for the PSF type -- by a joint fit on the three optical $g$-, $r$-, and $z$-band. 
The potential measurement bias are modeled with \citep{hildebrandt2012cfhtlens,miller2013bayesian,hildebrandt2017kids},
$$\gamma^{\mathrm{obs}}=(1+m) \gamma^{\text {true }}+c\ .$$
The additive bias $c$ and the multiplicative bias $m$ is expected to come from residuals in the anisotropic PSF correction, measurement method, blending and crowding \citep{mandelbaum2015great3,2019A&A...627A..59E}. 
This calibration is obtained by comparing with Canada–France–Hawaii Telescope (CFHT) Stripe 82 observed galaxies and Obiwan simulated galaxies \citep{phriksee2020weak,kong2020removing}.
For both DECaLS galaxies of LS DR8 and DR9 , we employ the photometric redshift based on \cite{zhou2021clustering}, which is estimated using the $g, r$, and $z$ optical bands from DECaLS and $W1$ and $W2$ infrared bands from WISE (Wide-field Infrared Survey Explorer, \citealt{wright2010wide}).

\subsection{Reconstruction}

To reconstruct the convergence map we select DECaLS galaxies of LS DR9 with $0.9\leq z_p\leq1.2$. Here $z_p$ is the best-fit value of the photometric redshift of a galaxy. 
For each bands we select galaxies with magnitude $m_g\leq23.5,m_r\leq23.0,m_z\leq22.5$.
Here $m_g,m_r,m_z$ are the magnitudes in $g,r,z$ band.
The magnitude cut of each band is set 0.5 lower than the peak position of the galaxy number counts as a function of the magnitude (\reffig{fig:a,g}).
This serves as an approximation for the flux-limit selection.
Then we equally divide galaxies of each band into $N_F=4$ flux bins.
This yields 12 flux bins in total. 
We have other sets of choice of the magnitude cuts and $N_F$ for consistency tests, which is described in Sec \ref{sec:robustest}.  Table \ref{tab:sub-sample} shows a summary of these galaxy sub-samples. 

For each flux bin, we project the 3D galaxy number density distribution to 2D sky map along the line-of-sight in the $N_{\rm side}$ = 2048 resolution.
We then downgrade the pixelized map to $N_{\rm side}$ = 1024.
{In order to minimize the impact from the footprint-induced fake overdensity, }
we select areas pixels with an observed coverage fraction $f_{\beta}$ larger than a threshold $f^{\rm threshold}$, where $f_{\beta}$ is defined as
\begin{equation}
    f_{\beta} = \frac{\sum\limits_{i=1}^{16}\mathcal{M}_{i}}{\sum\limits_{i}}
    \label{eq:fi} \ .
\end{equation}
Here $\mathcal{M}_i$ is the survey mask in $N_{\rm side}=2048$ map, which is created from the random catalogs by selecting random point with exposure time in all three bands greater than zero. For $f_\beta \geq f^{\rm threshold}$, we assign to each pixel $\beta$ its coverage fraction $f_\beta$. 
We use this value as a weight in the clustering measurements. 
We consider three choices of $f^{\rm threshold}$ later for consistency tests but fiducial one is 0.9.
After this selection, we get the galaxy overdensity for the 12 flux bins,
\be
    \hat{\delta}^\beta_i = \frac{N^\beta_i / f^\beta}{\langle N^\beta_i / f^\beta\rangle} -1 \label{eq:overdensity} \ .
\ee
Here $N_i^\beta$ is the galaxy number count of $i$-th flux bin in the $\beta$-th pixel. The average $\langle \cdots \rangle$ is over all pixels with $f\geq f^{\rm threshold}$.

We then apply the $\kappa$ estimator described in Section \ref{sec:method2} to the galaxy overdeisity maps. \reffig{fig:a,g} shows the galaxy number count as a function of magnitude and the estimated $\alpha$ and $g$ factors for each magnitude bin.
The numerical calculation of $\alpha$ of the flux bin  $F \in [f_1,f_2]$ is conducted by
\ba
\alpha=&\frac{\log \left(\int_{f_1 /(1+2 \delta_F)}^{f_2 /(1+2 \delta_F)} n(F) d F\right)-\log \left(\int_{f_1(1+2 \delta_F)}^{f_2 (1+2 \delta_F)} n(F) d F\right)}{2\log (1+2 \delta_F)}
\label{eq:a=dlogn_dlogf}
\ea
We find that the numerical calculation converges for $|\delta_F|<0.1 $ and we take the result from $\delta_F= 0.02$.

\begin{figure*}
\includegraphics[width=0.65\textwidth]{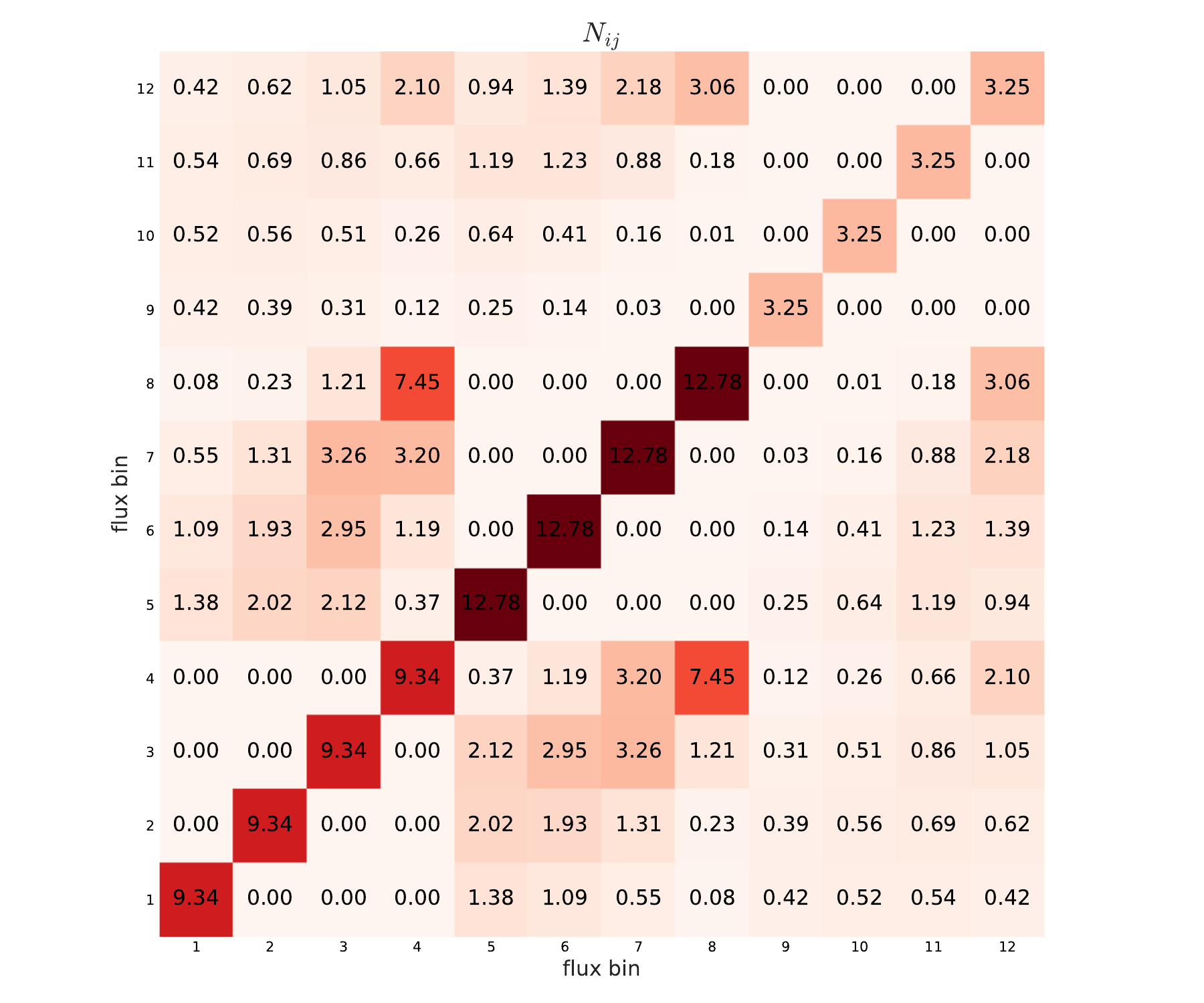}
\caption{The matrix $N_{ij}\equiv\frac{\bar{n}_{ij}}{\bar{n}_i\bar{n}_j}$ defined from the galaxy number density of the 12 flux bins/sub-samples.
The diagonal elements $N_{ii}\equiv\frac{1}{\bar{n}_i}$ are measured from the average number density of each flux bin (the fourth column in Table \ref{tab:sub-sample}). 
The off-diagonal elements $N_{ij}$ are measured from the overlap between the sub-samples in terms of the number of galaxies they contain (See the discussion in the caption of Table \ref{tab:sub-sample}).
This overlap arises form the fact that the relative ranking of galaxy brightness varies in different photometry  bands. 
 }
\label{fig:nij}
\end{figure*}

\begin{figure} 
\includegraphics[width=0.95\columnwidth]{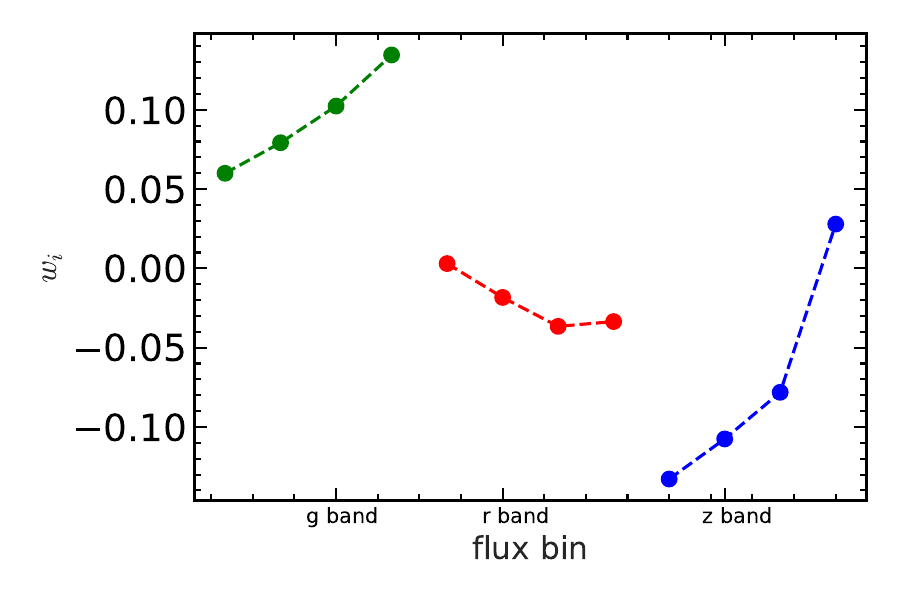}
\caption{The results of the weights calculated by Eq.\eqref{eq:weights} for each flux bin.
The estimator for the lensing convergence is obtained by calculating the weighted average of the galaxy number overdensity.
\label{fig:wi}}
\end{figure}

The results of the  $N_{ij}\equiv\frac{\bar{n}_{ij}}{\bar{n}_i\bar{n}_j}$ are shown in \reffig{fig:nij}.
Applying Eq.\eqref{eq:weights} yields the $w_i$'s for each flux bin, which is shown in \reffig{fig:wi}.
We then get the reconstructed convergence map sampled on HEALPix pixels by a weighted sum over the galaxy overdensity maps (Eq.\eqref{eq:linear combination}).
We select pixels with $f\geq f^{\rm threshold}$ for the cross correlation measurement.

\subsection{Imaging systematics}
\label{sec:imgweight}

Observation conditions, such as stellar contamination, Galactic extinction, sky brightness, seeing, and airmass, introduce spurious fluctuations in the observed galaxy density \citep[e.g.][]{ho2012clustering,10.1093/mnras/stv2103}.
Therefore, our reconstruction method based on galaxy density are potentially biased by these imaging systematics.
{One approach to mitigate imaging systematics is to weight the pixelized surface densities, ensuring that the weighted samples exhibit minimal dependence on imaging properties. 
Linear (or -quadratic) regression is commonly adopted to model the dependence of observed surface densities on various imaging maps.
But they may fail to capture higher-order dependencies on imaging features in strongly contaminated regions.
Recent studies have begun to employ machine learning algorithms to capture the higher-order dependencies,
such as Random Forest (RF) and Neural Network (NN) \citep[e.g.][]{10.1093/mnras/staa1231,10.1093/mnras/stab298}.\\
\indent
In this work, we apply the Random Forest (RF) mitigation \footnote{https://github.com/echaussidon/regressis} method developed by \citep[][]{10.1093/mnras/stab3252}, 
following the precedure described in \cite{HaojieXu2022UsingAT}.
It outperforms linear or quadratic models and achieves similar mitigation results at a significantly lower computational expense compared to Neural Network approach \cite{HaojieXu2022UsingAT}.
Applying the RF, a weight factor $w_{\mathrm{pix}, i}$ for each valid pixel $i$ will be returned. 
The imaging systematics can be reduced by weighting the 
reconstructed convergence map $\kappa_{\text {GRID }}$
according to their pixel weights. }
\begin{figure*} 
\centering
\includegraphics[width=1.9\columnwidth]{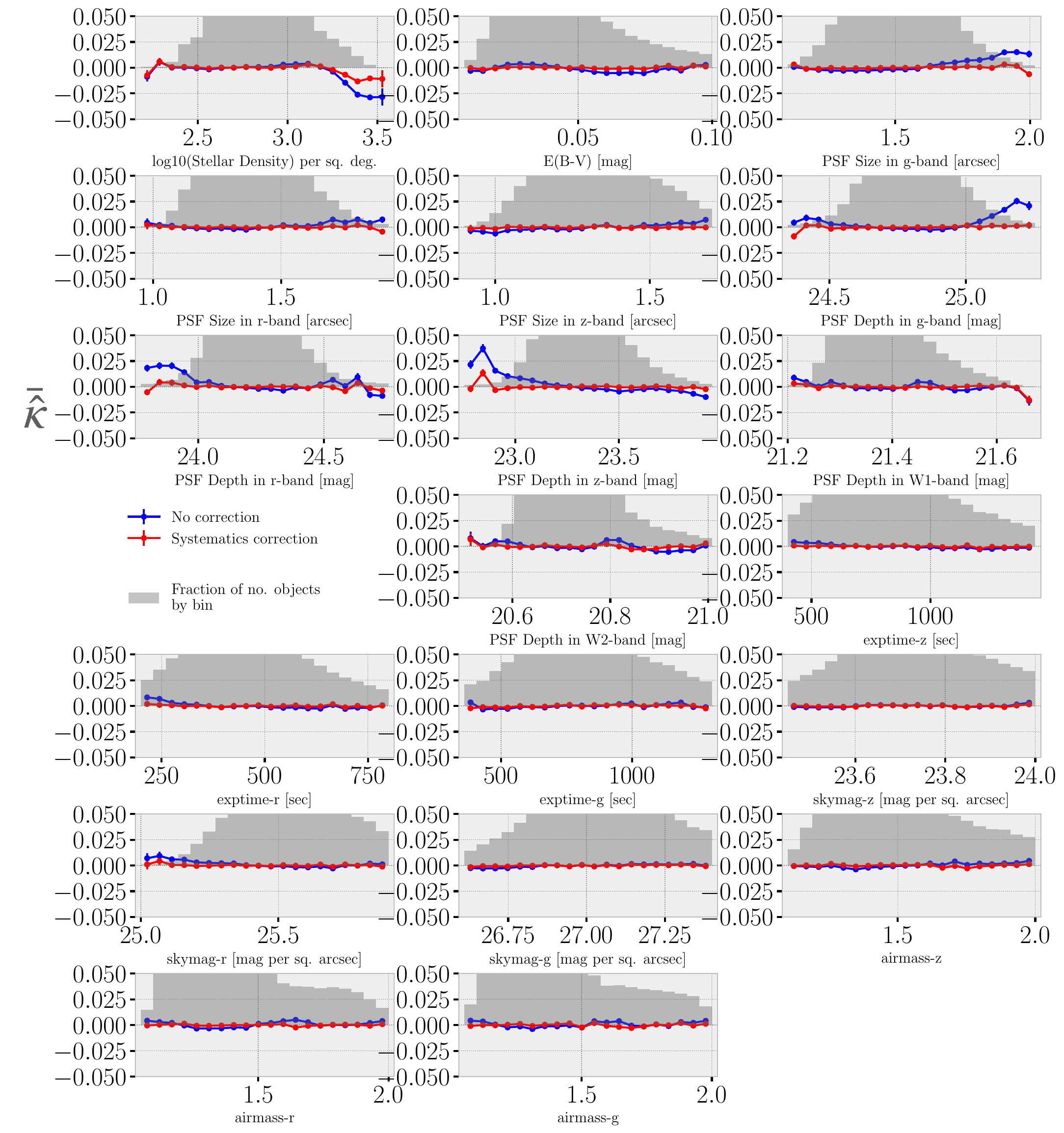}
\caption{Average value of the reconstructed convergence field as a function of 19 input imaging maps. The blue/red lines show the relative density before/after the systematics correction. The histogram is the fraction of pixels in bins of imaging properties, which is used to estimate the errors as the standard deviation of the convergence field in bins. After mitigating the imaging systematics, the convergence field show little fluctuation with respective to imaging properties. 
\label{fig:imagsys}}
\end{figure*}

\begin{figure*}
\centering
\includegraphics[width=2\columnwidth]{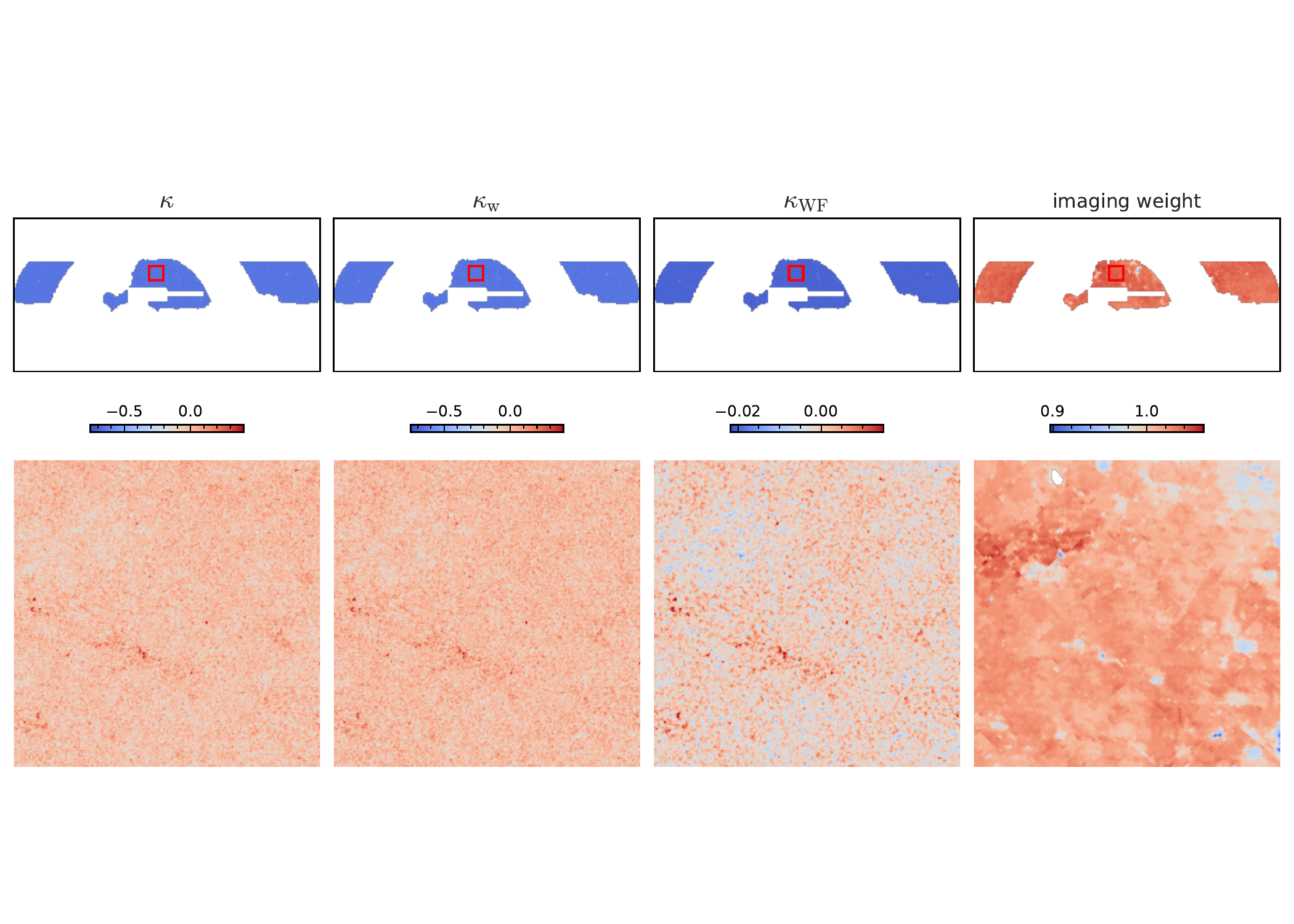}
\vspace{-2cm}
\caption[short]{The reconstructed lensing convergence map $\kappa_{\text {GRID }}$ sampled on HEALPix pixels with $N_{\rm side}$ = 1024. 
From left to right, the panels show the map before the imaging systematics mitigation, after the mitigation, and after the mitigation and application of the Wiener filter.
The raw convergence map is noise-dominated. 
After the Wiener filtering, the noise is significantly  reduced, and the large scale structures are more prominent. 
The lower panels correspond to zoomed-in views of the areas indicated by the red box in the upper panels.
The influence of the imaging systematics mitigation is not visually apparent, so we also show the imaging weight map in the rightmost panel.
The imaging weights help to increase the signal-to-noise ratio of the weak lensing signal, shown in the cross-correlation analysis.
 }
\label{fig:kappamap}
\end{figure*}

\reffig{fig:imagsys} shows the distribution of the reconstructed convergence field, before and after the mitigation, as a function of 19 input imaging properties. If the convergence field is independent of imaging properties, one would expect the mean value of $\hat\kappa$ in bins of imaging properties amounts to the global mean. We can see that the reconstructed field suffers some imaging contamination at percent level. After the mitigation, the corrected density is almost flat for all imaging properties.
Mitigating the imaging systematics enhances the detection significance of the weak lensing signal, which will be presented in Section \ref{sec:result4}.
After mitigating the imaging systematics, we get the final convergence map $\kappa_{\text {GRID }}$.
\reffig{fig:kappamap} shows the reconstructed convergence map before and after the imaging systematics mitigation. 
The Weiner filtered convergence map is also presented in \reffig{fig:kappamap}. The details of the Weiner filtering procedure are presented in the Appendix.

\begin{figure} 
\centering
\includegraphics[width=0.9\columnwidth]{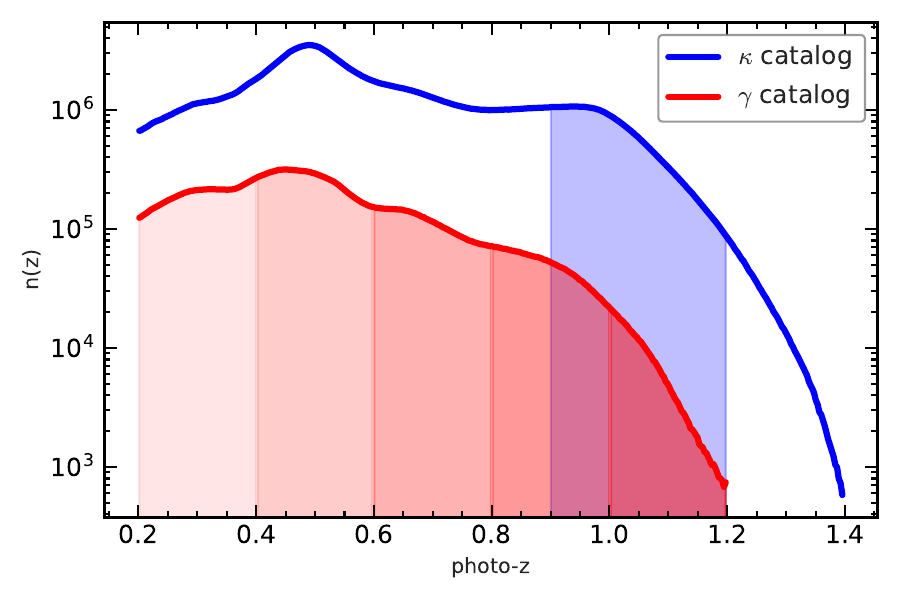}
\caption{The photometric redshift distributions of the DR9 galaxies used for lensing construction (blue line) and the DR8 galaxies with shear measurement (red line). 
The blue shadow regions indicate the photo-$z$ ranges utilized for lensing reconstruction in our study. 
The main results presented in the paper are based on the lensing convergence reconstructed from galaxies falling within the photo-$z$ range of $0.9<z_\kappa<1.2$.
The red shadow regions represent the photo-$z$ bins of the shear catalog, defined for cross-correlation analysis.
\label{fig:nz}}
\end{figure}

\subsection{Cross correlation measurement}\label{sec:tech}
To access the quality of the reconstructed map, we measure its cross-correlation with the DECaLS cosmic shear catalogs of LS DR8.
The two catalogs have an overlap in their footprints (Fig. \ref{fig:footprint}) of approximately $8365\deg^2$.
We select the shear galaxies with $0.2<z_{\rm p}<1.2$, which yields $\sim$35 million galaxy shear samples. 
The shear galaxies are then divided into nine photo-$z$ bins with edges at 0.2, 0.4, 0.6, 0.8, 1.0 and 1.2.
For each photo-$z$ bin, we estimate the convergence-shear cross-correlation function by
\be
\hat{\xi}^{\kappa\gamma}(\theta)=\langle \kappa_{\text {GRID }}\gamma_t\rangle_\theta/(1+\bar{m})\ .
\ee
Here the average $\langle \cdots \rangle_\theta$ is over all pixel-galaxy pairs within angular separation $\theta$, and $\bar{m}$ is the average shear multiplicative bias.
The calculations are done using the public available code TreeCorr\footnote{https://github.com/rmjarvis/TreeCorr}. We use $N=100$ Jackknife patches to estimate the covariance matrix
\be
\label{eq:covariance}
    \textbf{Cov}_{\alpha\beta} = \frac {N-1}{N} \sum\limits_{n=1}^{N=100} 
    [(\hat{\xi}^{\kappa\gamma}_n(\theta_\alpha) - \overline{\xi}^{\kappa\gamma}(\theta_\alpha))
    \times (\hat{\xi}^{\kappa\gamma}_n(\theta_\beta) - \overline{\xi}^{\kappa\gamma}(\theta_\beta)) ] \ .
\ee
Here $\overline{\xi}^{\kappa\gamma}$ is the average over the 100 patches. 
For each photo-$z$ bin, the data vector size is 9, and we rescale the covariance matrix following \cite{percival2014clustering,Wang_Zhao_Zhao_Philcox_Alam_Tamone_deMattia_Ross_Raichoor_Burtin2020} to get an unbiased estimation. 
\begin{figure*} 
\centering
\includegraphics[width=1.9\columnwidth]{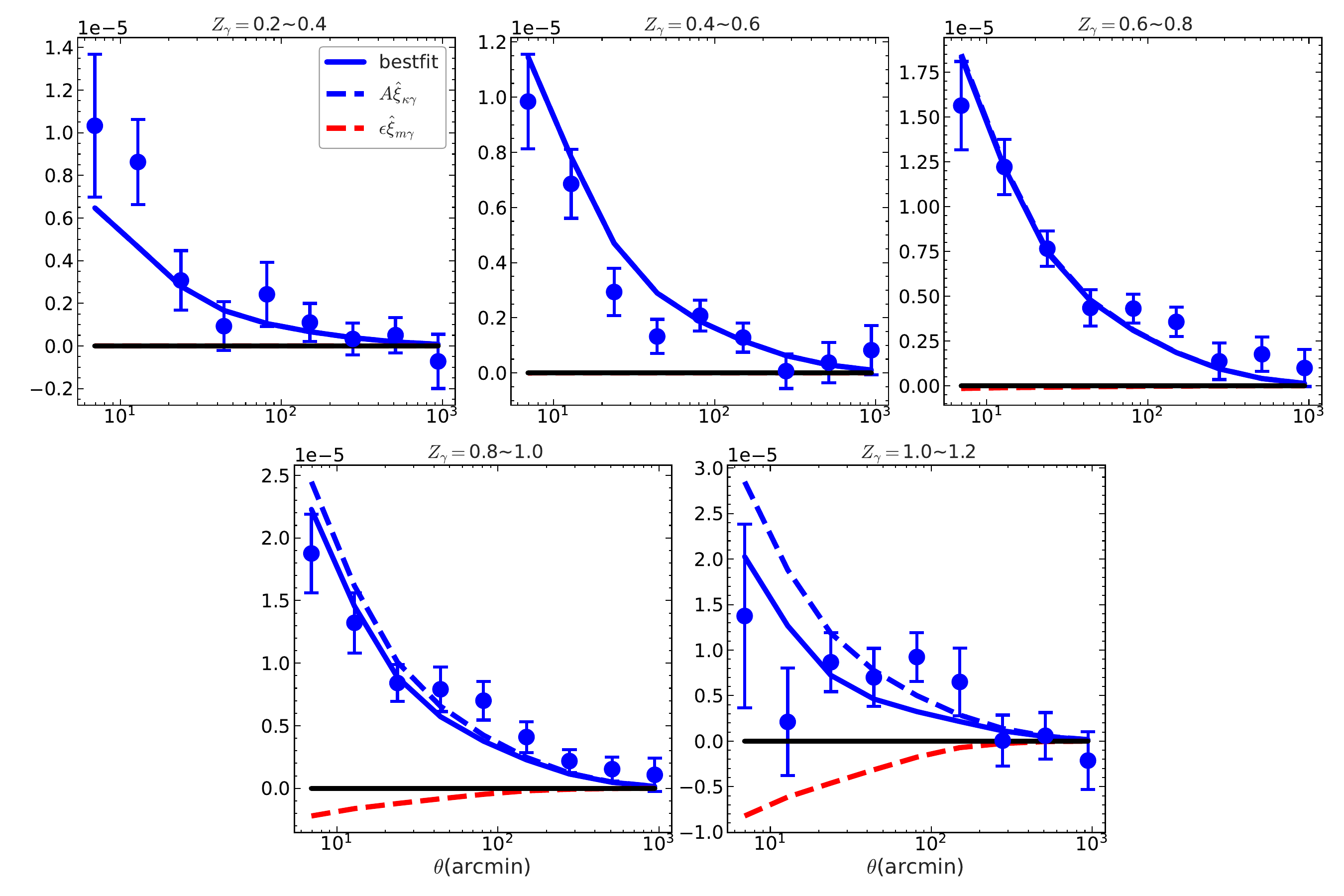}
\caption{Measured cross correlation function between the reconstructed lensing convergence and the shear, over the five photo-$z$ bins of the shear.
The blue solid lines represent the best-fit to the model in Eq.\eqref{eq:xikg_decom}.
The dashed lines denote the two components of the best-fit model, the signal term $A\xi^{\kappa\gamma}$ and the intrinsic clustering term $\epsilon\xi^{m\gamma}$.
The parameters $A$ and $\epsilon$ are defined to quantify the quality of the lensing reconstruction (Eq.\eqref{eq:k+em}).
They are kept same for all the photo-$z$ bins of the shear during the fitting.
The influence of selecting different bins of the shear is presented in the \ref{sec:robust_zgamma}.
For clarity we only show the results after the imaging systematics mitigation. 
The comparison of the fitting before and after the mitigation is summarized in Table \ref{tab:robust}.
\label{fig:xi}}
\end{figure*}

\begin{figure} 
\centering
\includegraphics[width=0.95\columnwidth]{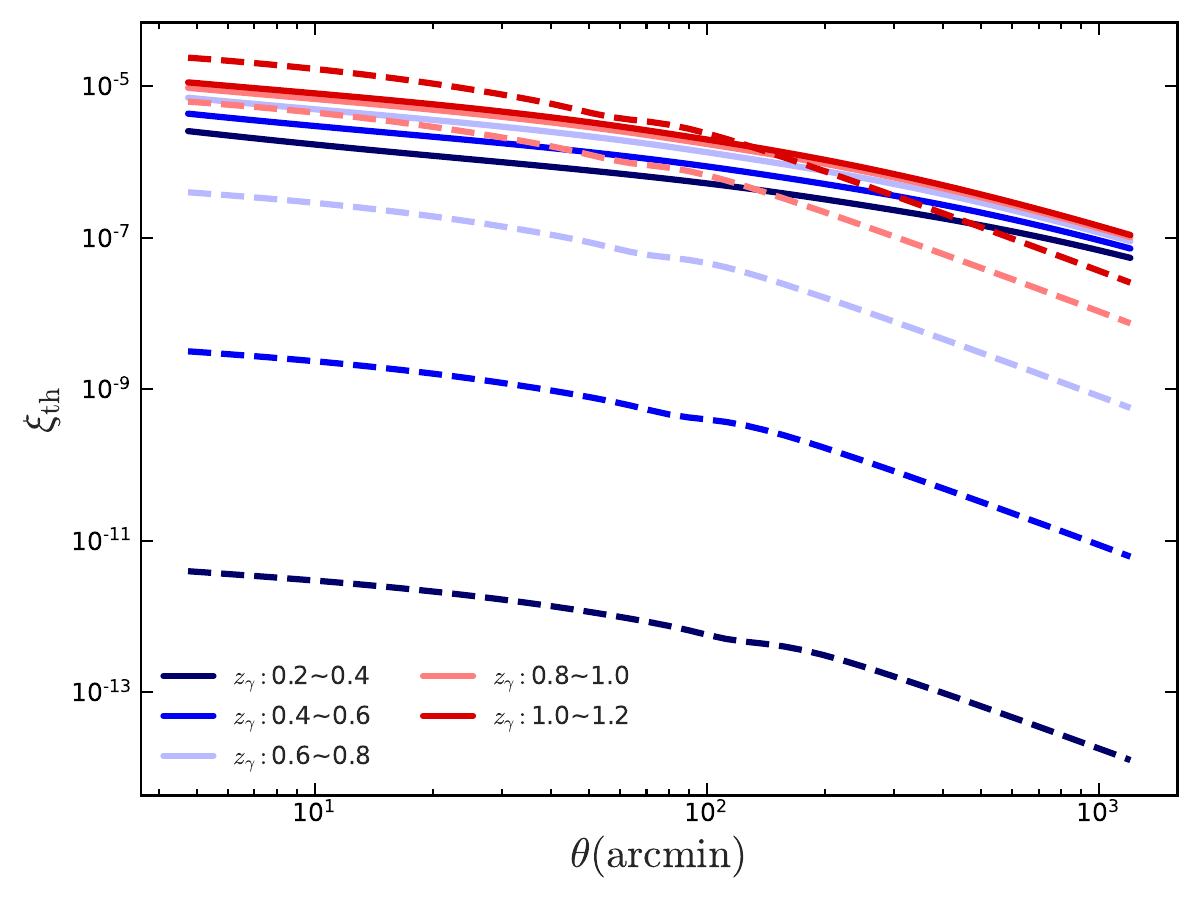}
\caption{The theoretical cross correlation templates, under the Planck 2018 cosmology. 
Different colors are used to distinguish the results of different photo-$z$ bins of the shear.
The signal term $\xi^{\kappa\gamma}$ are represented by the solid lines. 
The dashed lines represent the intrinsic clustering term $\xi^{m\gamma}$.
The difference in their shape and redshift dependence allows us to distinguish between the two terms.
Theoretically, $\xi^{m\gamma} = 0$ if the redshift of the shear is lower than that of the galaxies. 
However we use photometric redshifts whose true redshifts scatter to higher photo-$z$ bins,
so the intrinsic clustering term is non-zero and decreases drommatially with $z_\gamma$, for a Gaussian photo-$z$ error PDF with $\sigma_z=0.05$.
The photo-$z$ models will impact the theoretical templates.
We conducted tests using different assumed values of $\sigma_z$ and found that they do not affect the results of this work.
\label{fig:xith}}
\end{figure}

For the theoretical calculation of $\xi^{\kappa\gamma}(\theta)$ and $\xi^{m\gamma}(\theta)$, we apply the Core Cosmology Library (CCL, \citealt{pyccl}) and use  HaloFit \citep{Smith2003, Takahashi2012, Mead2015} to calculate the nonlinear matter power spectrum.
The galaxy redshift distribution $n(z)$ are calculated for each tomographic bin, combining the photo-$z$ distribution and a Gaussian photo-$z$ error PDF with $\sigma_z=0.05$\footnote{We conducted tests using different assumed values of $\sigma_z$ and found that they do not affect the  analysis results.}.
\reffig{fig:nz} displays the photo-$z$ distribution of the DR9 galaxies used for lensing construction and the DR8 shear catalog.
The theoretical results of $\xi^{\kappa\gamma}(\theta)$ and $\xi^{m\gamma}(\theta)$ are shown in \reffig{fig:xith}. 
The difference in their shape and redshift dependence allow us to distinguish between the two terms.


\section{Results}\label{sec:result4}
The results of the cross correlation are show in \reffig{fig:xi}.
Significant cross-correlation signals are obtained. We fit the measured cross correlation against the theoretical model (Eq.\eqref{eq:xikg_decom}) and constrain the two free parameters.
The constraints on $A$ can reveal the potential biases in the lensing reconstruction or in the cross correlation analysis, if its deviation from one is observed. The constraints on $\epsilon$ indicate the level of systematic errors.
The $\chi^2$ for the fitting can be approximated by
\ba
\label{eq:chi_1}
    \chi^2 \approx \sum\limits_{j\alpha\beta}
    \big[\hat{\xi}^{\kappa\gamma}_j(\theta_\alpha)-{\xi}_{j,\rm th}(\theta_\alpha)\big]
    \textbf{Cov}_{j\ \alpha\beta}^{-1}\big[\hat{\xi}^{\kappa\gamma}_j(\theta_\beta)-{\xi}_{j,\rm th}(\theta_\beta)\big]  \ .
\ea
Here $\hat{\xi}^{\kappa\gamma}_j$ and ${\xi}_{j,\rm th}$ denote the measurement and the model in the $j$-th photo-$z$ bin of shear, respectively. \textbf{Cov} is the data covariance matrix.\footnote{Here we have neglected correlations between  $\xi_j^{\kappa\gamma}$ of different shear redshifts ($j_{1,2}$) arising from the four-point correlation ($\langle \gamma_{j1}\kappa\gamma_{j2}\kappa\rangle-\langle \gamma_{j1}\kappa\rangle \langle \gamma_{j2}\kappa\rangle)\sim \langle \gamma_{j1}\kappa\rangle \langle \gamma_{j2}\kappa\rangle+\langle \gamma_{j1}\gamma_{j2}\rangle\langle \kappa\kappa\rangle $. For the current data, is is negligible comparing to the shape measurement error in $\gamma$ and shot noise in $\kappa$. }
$A$ and $\epsilon$ are identical for each photo-$z$ bin and so the sum is over all the data points and photo-$z$ bins.

\begin{figure} 
\centering
\includegraphics[width=\columnwidth]{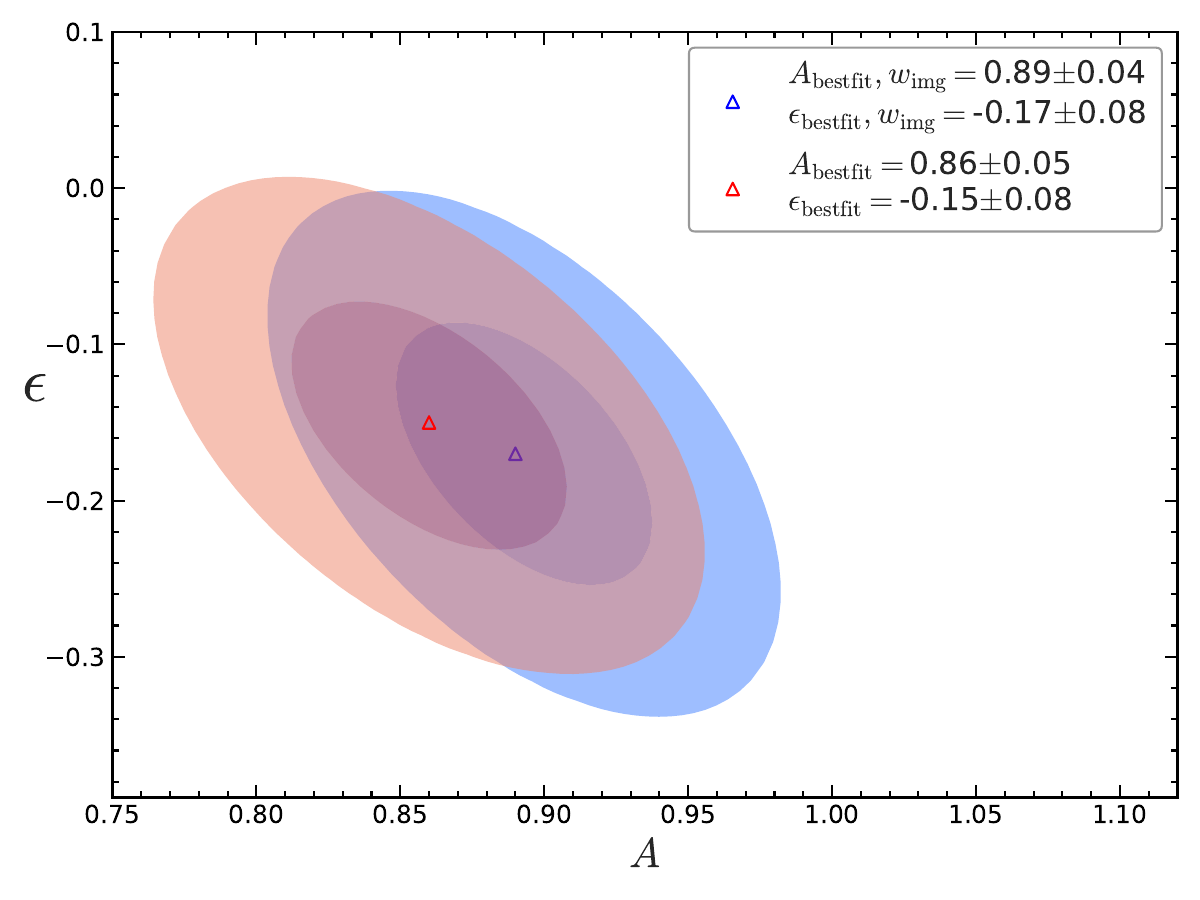}
\caption{Results of the constraints on $A$ and $\epsilon$ from the cross-correlation analysis.  
The red/blue contours represent the 1- and 2-$\sigma$ confidence regions before/after the imaging systematics mitigation.
The triangular markers indicate the best-fit values.
The results are consistent before and after the mitigation, and the imaging systematics mitigation enhances the detection significance of the weak lensing signal.
\label{fig:Ae}}
\end{figure}

The best fit is
\ba
\big(A^{\text{bestfit}}\ \epsilon^{\text{bestfit}}\big)^{\mathbf{T}}=\bm{F}^{-1}\bigg[\sum_j\bm{\xi}_j^{\mathbf{T}}\textbf{Cov}_j^{-1}\bm{\hat{\xi}}_j\bigg]\ ,
\ea
where
\begin{equation}
\bm{\xi}_j\equiv\begin{pmatrix}
\xi_j^{\kappa\gamma}(\theta_1) & \xi_j^{\kappa m}(\theta_1) \\ 
\xi_j^{\kappa\gamma}(\theta_2) & \xi_j^{\kappa m}(\theta_2)\\
\vdots & \vdots\\
\xi_j^{\kappa\gamma}(\theta_n) & \xi_j^{\kappa m}(\theta_n)
\end{pmatrix}; 
\hat{\bm{\xi}}_i\equiv
\begin{pmatrix}
\hat{\xi}_j^{\kappa\gamma}(\theta_1)  \\ 
\hat{\xi}_j^{\kappa\gamma}(\theta_2) \\
\vdots\\
\hat{\xi}_j^{\kappa\gamma}(\theta_n) 
\end{pmatrix}\ .
\end{equation}
The associated errors and covariance matrix of the constraints are given by $\bm{F}^{-1}$. $\bm{F}$ is the fisher matrix of $A$ and $\epsilon$, 
\ba
\bm{F} =& 
\begin{pmatrix} 
\sum\limits_{j\alpha\beta}\xi^{\kappa\gamma}_{j,\alpha}\textbf{Cov}_{j,\alpha\beta}^{-1}\xi^{\kappa\gamma}_{j,\beta} & \sum\limits_{j\alpha\beta}\xi^{\kappa\gamma}_{j,\alpha}\textbf{Cov}_{j\ \alpha\beta}^{-1}\xi^{\kappa m}_{j,\beta}\nonumber \\ 
\sum\limits_{j\alpha\beta}\xi^{\kappa\gamma}_{j,\alpha}\textbf{Cov}_{j\ \alpha\beta}^{-1}\xi^{\kappa m}_{j,\beta} & \sum\limits_{j\alpha\beta}\xi^{\kappa m}_{j,\alpha}\textbf{Cov}_{j\ \alpha\beta}^{-1}\xi^{\kappa m}_{j,\beta} \nonumber
\end{pmatrix}\
\ea
\reffig{fig:Ae} shows the results of the constraints.
According to $A$, we detect a convergence-shear cross-correlation signal with $S/N\equiv A/\sigma_A\simeq 22$. 
No significant deviation from one is detected in $A$, which indicates that multiplicative bias in the reconstructed convergence is minimal. 
According to $\epsilon$, the galaxy intrinsic clustering is suppressed by a factor by a factor $\sim 10$, to a level of $-0.17\pm 0.08$.
The best-fit to the cross-correlation measurements is shown in \reffig{fig:xi}, and the two components $A\xi^{\kappa\gamma}$ and $\epsilon\xi^{m\gamma}$ are also highlighted in the figure. 
The intrinsic clustering term $\epsilon\xi^{m\gamma}$ is subdominant for all cases.  


Table \ref{tab:robust} lists the results of $\chi^2_{\rm min}$ and ${\chi^2_{\rm null}}$to demonstrate the goodness of fit to the model, where the $\chi^2_{\rm null}$ and $\chi^2_{\rm min}$ is  defined by
\ba
\label{eq:chi_1}
    \chi^2_{\rm null} = \sum\limits_{j\alpha\beta}
    \big[\hat{\xi}^{\kappa\gamma}_j(\theta_\alpha)\big]
    \textbf{Cov}_{j\ \alpha\beta}^{-1} 
    \times\big[\hat{\xi}^{\kappa\gamma}_j(\theta_\beta)\big] \ ,
\ea
\ba
\label{eq:chi_1}
    \chi^2_{\rm min} = \sum\limits_{j\alpha\beta}
    \big[\hat{\xi}^{\kappa\gamma}_j(\theta_\alpha)
    -{\xi}_{j,\rm th}^{\rm bestfit}(\theta_\alpha)
    \big]
    \textbf{Cov}_{j\ \alpha\beta}^{-1} \nonumber\\
    \times\big[\hat{\xi}^{\kappa\gamma}_j(\theta_\beta)
    -{\xi}_{j,\rm th}^{\rm bestfit}(\theta_\alpha)
    \big] \ ,
\ea
where
\be
\xi_{\rm th}^{\rm bestfit}=
A^{\rm bestfit}\xi^{\kappa\gamma}+\epsilon^{\rm bestfit}\xi^{m\gamma}\ .
\ee
The degree-of-freedom (d.o.f) of the fitting is $43$ and the $\chi^2_{\rm min}\simeq48$.
Therefore, the two parameter fitting returns reasonable $\chi^2_{\rm min}$/d.o.f. $\sim 1$. 
This means that shape of the measured cross-correlation agrees with the model prediction, and provides support that the detected signal is cosmological in origin.
With 45 non-independent data points and the full covariance, $\chi^2_{\rm null} = 550.2$ and  $\chi^2_{\min} = 48.4$ with respect to the null and to the model, respectively. 
According to the data-driven signal-to-noise ratio $\sqrt{\chi^2_{\rm null}}$, we get $23.5 \sigma$ detection of a non-zero cross-correlation signal. 
According to the fitting-driven signal-to-noise ratio $\sqrt{\chi^2_{\rm null}-\chi^2_{\rm min}}$, the significance is $\sim 22.4\sigma$. 
We compared the results before and after the imaging systematics mitigation in Table \ref{tab:robust}.
It shows that the fitting results are consistent, and after the mitigation, both $\sqrt{\chi^2_{\rm null}-\chi^2_{\rm min}}$ and $A/\sigma_A$ exhibit an increase from 19.9 to 22.4 and 17.2 to 22.2, respectively.

\subsection{Internal test}\label{sec:robustest}
We test the impact of several factors in the analysis, which are summarized in Table \ref{tab:robust}.
First, we compare the results for $f^{\rm threshold}=0.5, 0.7,0.9$.
Larger $f^{\rm threshold}$ indicates
more stringent
selection of the pixels of galaxy number overdensity map.  Table \ref{tab:robust} shows that the impact of $f^{\rm threshold}$ is small, and the fitting result is stable for different $f^{\rm threshold}$.

Second, we test the cases of $N_F=2, 3, 4$ and find that the constraints on $A$ and $\epsilon$ are stable for different $N_F$.
The $\epsilon=\Sigma_i w_i b_i$ depends on the galaxy biases in the selected flux bins and, consequently, on the number of flux bins $N_F$.
The $\epsilon$ is consistently to be $\sim -0.1$ for different $N_F$.
It indicates that our method of eliminating the galaxy intrinsic clustering is robust.

Third, we test the influence of
the magnitude cut on the results. 
We select galaxies which are 0.5 magnitude brighter than the baseline set. 
The first case involves selecting galaxies that are 0.5 magnitudes brighter across all three bands, denoted as $m_{g,r,z}-0.5$. 
The other three cases involve selecting galaxies that are 0.5 magnitudes brighter in each band individually, denoted as $m_{g}-0.5$, $m_{r}-0.5$ and $m_{z}-0.5$.
As shown in Table \ref{tab:robust}, the fitting results are stable for different magnitude cuts.
A brighter magnitude cut of 0.5 results in a decrease of $\sim$ 30$-$40\% in the number of galaxies for a particular photometry band.
Therefore, the S/N of the cross-correlation signal is reduced for a brighter magnitude cut.
The value of $\epsilon$ remains minimal in all cases and is even smaller than the baseline, reaching $\sim -0.01$.

In addition, Table \ref{tab:robust} presents the comparison of fitting results before and after mitigating the imaging systematics for all the tests investigated above. Across all cases, the fitting results are consistent and exhibit enhancements in the S/N after the mitigation, indicating the effectiveness and robustness of the mitigation procedure.

\subsection{Model selection}

\begin{table}
\centering
\begin{tabular}{llllll}
\hline
\hline
    {} & baseline & A=0 & A=1 & $\epsilon$=0 \\
$\chi^2_{\rm min}$  &    48.4 &  449.3 &  54.1 &          52.5 \\
AIC                 &    52.6 &  451.4 &  56.2 &          54.5 \\
\hline
\end{tabular}
\caption{The results of the Akaike information criterion (AIC) analysis for different models.
The baseline model is the one described by Eq. \eqref{eq:xikg_decom}. 
The other models fix $A=0$, $A=1$, or $\epsilon=0$. 
The model that fixes $\epsilon=0$ has the lowest AIC score, indicating that the residual galaxy clustering contamination in the $\hat{\kappa}$ map is insignificant. 
The model that fixes $A=1$ has AIC score lower than the baseline, suggesting the insignificance of the multiplicative bias in the overall amplitude of $\hat{\kappa}$.
The model that fixes $A=0$ has an AIC score that is $\sim 200$ higher than the baseline, ruling out null detection of weak lensing signal. }
\label{tab:AIC}
\end{table}

Performing an Akaike information criterion (AIC) analysis can provide valuable insights into which model is the most suitable for explaining the observed  data. 
In this part, we modify the theoretical template to facilitate the comparison of different models.
In addition to the baseline model described by Eq. \eqref{eq:xikg_decom}, we also consider alternative models which we keep either $A$ or $\epsilon$ fixed.
We compare four cases: the baseline model, fixing $A=0$, fixing $A=1$, and fixing $\epsilon=0$.
For each model we repeat the fitting process and then calculate the AIC, which can be expressed, to second-order, as:
$$
\mathrm{AIC}_{\mathrm{c}}=-2 \ln p\left(\lambda_{\text {bestfit }} \mid D\right)+2 N_\lambda+\frac{2 N_\lambda+1}{N_D-N_\lambda-1} .
$$
Here $N_\lambda$ is the number of parameters, $N_D$ is the number of data points, $p$ is the likelihood which is Gaussian in our case and $p\left(\lambda_{\text {bestfit }} \mid D\right)$ is the value for the best-fit parameters.
The "best" model has the smallest AIC. If another model's AIC is larger by 10 or more, this model should be ruled out. If the difference is less than 2, the two models can not be really distinguished.

Table \ref{tab:AIC} shows the results of $\chi^2_{\rm min}$ (the minimum $\chi^2$ value obtained during fitting) and AIC.
The baseline model has the lowest AIC score.
The model that fixes $A=0$ has an AIC score that is roughly 400 higher than the baseline, ruling out null detection of weak lensing signal.
The model that fixes $\epsilon=0$ is not ruled out by the data, with an AIC score that is only slightly higher ($\lesssim 2$) than the baseline.
It indicates that the residual galaxy clustering contamination in the reconstructed $\hat{\kappa}$ map is insignificant.

\begin{table*}
\begin{center}
\begin{tabular}{lllllll}
\hline\hline
{} &                             $A$ & $A_{\rm precition}$ &                      $\epsilon$ & $\chi^2_{\rm min}$ & $\sqrt{\chi^2_{\rm null}-\chi^2_{\rm min}}$ & $A/\sigma_A$ \\
\midrule
\hline
baseline                       &    0.89$\pm$0.04(0.86$\pm$0.05) &                1(1) &  -0.17$\pm$0.08(-0.15$\pm$0.08) &         48.4(52.3) &                                  22.4(19.9) &   22.2(17.2) \\
f=0.5                          &    0.87$\pm$0.05(0.83$\pm$0.05) &                1(1) &  -0.15$\pm$0.08(-0.18$\pm$0.08) &         57.2(55.0) &                                  21.2(18.8) &   17.4(16.6) \\
f=0.7                          &    0.87$\pm$0.05(0.85$\pm$0.05) &                1(1) &  -0.11$\pm$0.08(-0.19$\pm$0.09) &         56.6(58.4) &                                  20.8(18.9) &   17.4(17.0) \\
$N_F=2$                        &    0.90$\pm$0.05(0.88$\pm$0.05) &                1(1) &  -0.14$\pm$0.08(-0.20$\pm$0.09) &         51.1(51.3) &                                  21.2(20.1) &   18.0(17.6) \\
$N_F=3$                        &    0.86$\pm$0.04(0.88$\pm$0.05) &                1(1) &  -0.13$\pm$0.08(-0.21$\pm$0.08) &         52.4(49.6) &                                  21.8(20.1) &   21.5(17.6) \\
$m_{g,r,z}-0.5$                &    0.98$\pm$0.06(0.94$\pm$0.06) &                1(1) &   0.01$\pm$0.09(-0.01$\pm$0.10) &         51.5(49.2) &                                  19.4(17.6) &   16.3(15.7) \\
$m_g-0.5$                      &    0.83$\pm$0.05(0.82$\pm$0.05) &                1(1) &   0.00$\pm$0.08(-0.07$\pm$0.07) &         52.6(51.3) &                                  20.1(20.1) &   16.6(16.4) \\
$m_r-0.5$                      &    0.90$\pm$0.05(0.86$\pm$0.05) &                1(1) &  -0.13$\pm$0.08(-0.19$\pm$0.09) &         54.5(55.3) &                                  22.1(19.7) &   18.0(17.2) \\
$m_z-0.5$                      &    1.02$\pm$0.05(1.01$\pm$0.05) &                1(1) &  -0.06$\pm$0.10(-0.08$\pm$0.09) &         55.7(46.5) &                                  21.0(20.8) &   20.4(20.2) \\
\hline
\hline
$g_i+{\rm Err}_{5\%,i}$        &    0.85$\pm$0.04(0.85$\pm$0.04) &          0.99(0.99) &  -0.06$\pm$0.07(-0.15$\pm$0.08) &         48.9(47.3) &                                  24.4(21.8) &   21.2(21.2) \\
$g_i+{\rm Err}_{10\%,i}$       &    0.97$\pm$0.05(0.96$\pm$0.05) &          0.96(0.96) &  -0.07$\pm$0.09(-0.13$\pm$0.10) &         47.8(50.9) &                                  22.9(21.4) &   19.4(19.2) \\
$g_i+{\rm Err}_{15\%,i}$       &    0.78$\pm$0.05(0.76$\pm$0.06) &          0.75(0.75) &  -0.06$\pm$0.08(-0.12$\pm$0.08) &         59.2(65.4) &                                  18.4(15.7) &   15.6(12.7) \\
\hline
$\sum w_i = 1, \sum w_i g_i=0$ &    0.07$\pm$0.16(0.02$\pm$0.16) &          1.00(1.00) &    0.46$\pm$0.31(0.53$\pm$0.31) &         35.3(47.8) &                                    2.1(2.2) &     0.4(0.1) \\
\hline
\end{tabular}
\end{center}
    \caption{Summary of the cross-correlation analysis results for various tests.
    In the first part, we test the impact of $f^{\rm threshold}, N_{\rm F}$, the three bands magnitude cuts on the analysis.
    The fiducial/baseline sets are $f^{\rm threshold}=0.9, N_{\rm F}=4, m_g=23.4, m_r=22.8, m_z=22.2$. 
    The fitting results are stable for different $f^{\rm threshold}, N_{\rm F}$, and magnitude cuts.
    The second part tests the impact of the errors/biases in the magnification coefficient on the analysis.
    Three levels of errors/biases are tested, denoted as ${\rm Err}_{5\%,i}$, ${\rm Err}_{10\%,i}$, and ${\rm Err}_{15\%,i}$.
    Errors/biases in the magnification coefficient introduce biases in the overall amplitude of reconstructed $\hat{\kappa}$.
    They are indicated by the deviation of $A$ from one and are predicatable (denoted as $A_{\rm prediction}$) given the results of $w_i$ and $g_i$.
    The third part (the last row) involves constructing the intrinsic clustering ($\sum w_i = 1$) while suppressing the lensing magnification ($\sum w_i g_i=0$).
    $A$ is significantly suppressed and the residual is tiny ($A/\sigma_A\sim 0$), indicating the measurement of $g_i$ is reliable for this case.
    The d.o.f. = 43.  
    The main table shows the results after the mitigation, while in parentheses are the results before the mitigation.
    The S/N of the detection of the weak lensing signal is enhanced after the mitigation.
      }    \label{tab:robust}
\end{table*}

\begin{table*}
\begin{center}
\begin{tabular}{lllllll}
\hline\hline
{} &                             $A$ & $A_{\rm precition}$ &                      $\epsilon$ & $\chi^2_{\rm min}$ & $\sqrt{\chi^2_{\rm null}-\chi^2_{\rm min}}$ & $A/\sigma_A$ \\
\midrule
\hline
baseline                       &    2.55$\pm$0.25(2.59$\pm$0.28) &                1(1) &  -0.00$\pm$0.02(-0.01$\pm$0.02) &         26.6(29.4) &                                  17.5(15.8) &    10.2(9.2) \\
f=0.5                          &    2.56$\pm$0.25(2.69$\pm$0.31) &                1(1) &  -0.00$\pm$0.02(-0.02$\pm$0.03) &         22.6(30.9) &                                  17.6(14.5) &    10.2(8.7) \\
f=0.7                          &    2.61$\pm$0.23(2.69$\pm$0.31) &                1(1) &  -0.01$\pm$0.02(-0.02$\pm$0.03) &         23.4(30.2) &                                  17.7(14.4) &    11.3(8.7) \\
$N_F=2$                        &    2.61$\pm$0.25(2.64$\pm$0.30) &                1(1) &  -0.01$\pm$0.02(-0.01$\pm$0.03) &         27.0(31.0) &                                  17.7(15.6) &    10.4(8.8) \\
$N_F=3$                        &    2.54$\pm$0.24(2.69$\pm$0.30) &                1(1) &  -0.00$\pm$0.02(-0.03$\pm$0.03) &         25.0(26.0) &                                  17.6(14.4) &    10.6(9.0) \\
$m_{g,r,z}-0.5$                &    2.62$\pm$0.30(2.70$\pm$0.34) &                1(1) &    0.03$\pm$0.03(0.02$\pm$0.03) &         23.9(31.2) &                                  16.4(14.2) &     8.7(7.9) \\
$m_g-0.5$                      &    2.35$\pm$0.25(2.41$\pm$0.30) &                1(1) &    0.03$\pm$0.02(0.02$\pm$0.03) &         28.4(35.4) &                                  17.7(14.9) &     9.4(8.0) \\
$m_r-0.5$                      &    2.43$\pm$0.25(2.52$\pm$0.29) &                1(1) &   0.00$\pm$0.02(-0.02$\pm$0.03) &         21.2(29.8) &                                  16.9(14.2) &     9.7(8.7) \\
$m_z-0.5$                      &    2.78$\pm$0.27(2.82$\pm$0.33) &                1(1) &  -0.03$\pm$0.03(-0.03$\pm$0.03) &         17.2(23.4) &                                  15.5(13.3) &    10.3(8.6) \\
\hline
$g_i+{\rm Err}_{5\%,i}$        &    2.30$\pm$0.23(2.37$\pm$0.27) &          0.92(0.92) &  -0.02$\pm$0.02(-0.03$\pm$0.02) &         21.7(31.8) &                                  16.6(13.4) &    10.0(8.8) \\
$g_i+{\rm Err}_{10\%,i}$       &    2.49$\pm$0.24(2.52$\pm$0.32) &          0.80(0.80) &  -0.05$\pm$0.02(-0.06$\pm$0.03) &         16.3(17.7) &                                  14.2(10.8) &    10.4(7.9) \\
$g_i+{\rm Err}_{15\%,i}$       &    1.77$\pm$0.19(1.87$\pm$0.21) &          0.47(0.47) &    0.02$\pm$0.02(0.01$\pm$0.02) &         33.0(38.8) &                                  18.5(17.1) &     9.3(8.9) \\
\hline
$\sum w_i = 1, \sum w_i g_i=0$ &  -3.11$\pm$1.00(-3.42$\pm$1.10) &                0(0) &    1.29$\pm$0.09(1.35$\pm$0.09) &         46.5(59.6) &                                  22.2(24.5) &   -3.1(-3.1) \\
\hline
\end{tabular}
\end{center}
    \caption{Same as Table \ref{tab:robust}, but for lensing convergence constructed at $0.5 < z_\kappa < 0.8$.
    The convergence-shear cross correlation is detected at $S/N \sim 10$ with the intrinsic clustering well suppressed.
    The results are stable for different $f^{\rm threshold}, N_{\rm F}$, and magnitude cuts.
    However there are three main differences from the results of $0.9 < z_\kappa < 1.2$.
    First, the lensing amplitude $A$ of the baseline has bias of a factor of $\sim 2$.
    Second, the biases in $A$ induced by the errors/biases in $g$ differs from the prediction $A_{\rm prediction}$.
    Third, the intrinsic clustering construction ($\sum w_i = 1$) while suppressing the lensing magnification ($\sum w_i g_i=0$) returns a significant residual in $A$.
    All the three anomalies implies that the logarithmic slope of the luminosity function should not be a good approximation for the magnification coefficient for $ 0.5 < z_\kappa < 0.8$.    
      }    \label{tab:robust5-8}
\end{table*}

\subsection{Impact of the shear redshifts}\label{sec:robust_zgamma}
\begin{figure}
\centering
\includegraphics[width=\columnwidth]{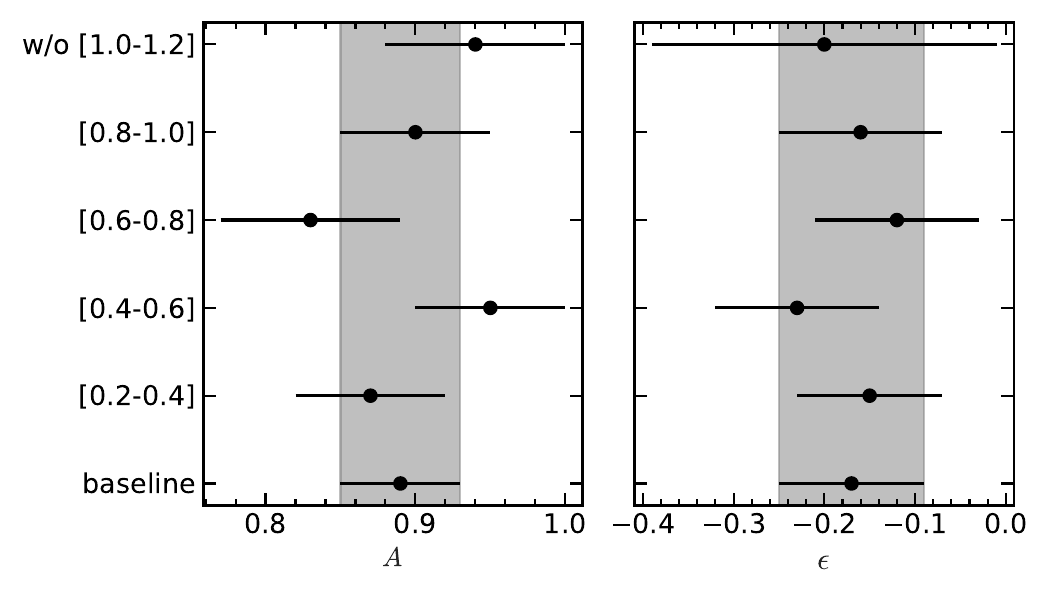}
\caption{Impact of the shear redshifts on the constraints of $A$ (the left panel) and $\epsilon$ (the right panel).
We drop one shear redshift bin at a time, and perform the cross correlation analysis and constrain the parameters with the remaining bins. 
The y-axis denote the shear redshift bin that is dropped.
The shadow regions shows the $1 \sigma$ error of the constraints for the baseline set (i.e., all the shear bins are included).
The constraining results are consistent with the baseline set for all the cases, 
which indicates the robustness of the cross-correction analysis against different shear samples.
}
\label{fig:Ae_dropZg}
\end{figure}

The parameters $A$ and $\epsilon$ characterize the reconstructed lensing convergence and are therefore independent of the shear catalog utilized. Consequently, the constraints on $A$ and $\epsilon$ are expected to remain consistent when different shear catalogs are employed for cross-correlation analysis. In this part, we investigate the impact of selecting different shear catalogs on the analysis, by systematically excluding one shear photo-$z$ bin at a time. We then perform cross-correlation analysis and constrain $A$ and $\epsilon$ using the remaining photo-$z$ bins / shear catalogs. The results of the test are presented in Fig. \ref{fig:Ae_dropZg}. For all the investigated cases, the constraints on $A$ and $\epsilon$ align with the baseline set within the $1\sigma$ error range,
indicating the robustness of the analysis against different shear samples.


\subsection{Impact of error in $g$}\label{sec:robust_g}
\begin{figure}
\centering
\includegraphics[width=\columnwidth]{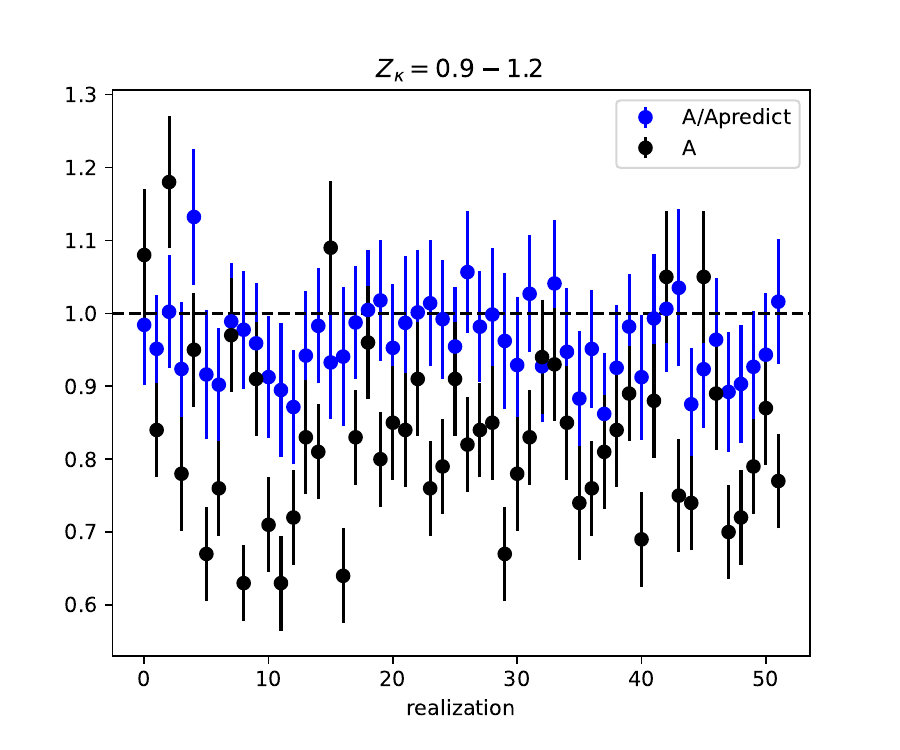}
\caption{   Comparison of the constraints on $A$ and the prediction $A_{\rm prediction}$ for the 50 realizations of the Gaussian error in $g$ at the $15\%$ level.
The black data points denote the results of the constraints on $A$,
with errorbars representing the $1\sigma$ error in the constraints.
The blue data points denote the ratio of $A$ to $A_{\rm prediction}$,
whose errors are propagated from the errors in $A$.
For different realizations, the errors in $g$ induces biases in the reconstructed $\hat{\kappa}^b$,
which makes $A$ deviate from $1$.
The biases agree well with the prediction, i.e., $A/A_{\rm prediction}\sim 1$.
The prediction is based on the assumption that $g_i^{\rm true}$ can be approximated by the logarithmic slope of the luminosity function. 
If this approximation does not hold, the bias in A may differ from the predicted value $A_{\rm prediction}$. 
This occurs for $ 0.5 < z_\kappa < 0.8$ (see Table \ref{tab:robust5-8} and \reffig{fig:A_Apredict_z5-8}), 
suggesting that, if no other systematics are present, 
the systematic errors in $g$ should be the source of the bias in $A$ for the reconstructed lensing convergence for $ 0.5 < z_\kappa < 0.8$.
}
\label{fig:A_Apredict_z9-12}
\end{figure}

{The magnification coefficient $g$ is a key parameter in the lensing reconstruction.
The error in $g$ can lead to biases in the reconstructed lensing convergence.
This bias can propagated the cross-correlation analysis and affect the constraints on $A$.
To demonstrate this point, for each flux bin $i$, we introduce an error in $g$ by adding Gaussian noise to the magnification coefficient, $g_i \rightarrow g_i + E_{i,\rm sys}$. 
Here, $E_{i,\rm sys}$ is Gaussian noise with zero mean and standard deviations of $5\%$, $10\%$, and $15\%$ of $g_i$, 
to mimic three levels of \red{observational statistical} errors.
With the biased $g^b_i$, we contruct the lensing reconstruction with the same procedure as described in Sec. \ref{sec:method2}.
And it returns the biased lensing convergence $\hat{\kappa}^b$ and the weights $w_i^b$ for each flux bin.
The bias in $\hat{\kappa}^b$ can then be predicted from the results of $w_i^b$ and $g_i$.
From Eq. \eqref{eq:linear combination}, the estimator of lensing convergence with $w_i^b$ becomes
\be\label{eq:linear combination}
\hat{\kappa}^b=\sum_{i} w^b_{i}\delta_{i}^{\rm L}\ . 
\ee 
With Eq. \eqref{delta} the above equation reads
\be
\hat{\kappa}^b= \sum_{i} w^b_{i}g^{\rm true}_i \kappa + \cdots \ .
\ee
Where $\cdots$ denotes the other terms (e.g., the intrinsic clustering, the galaxy stochasticity) which are not relevant for the current discussion.
$g^{\rm true}_i$ is the true magnification coefficient, which we can approximate by $g_i$.
Therefore, the prediction for the biased lensing convergence is
\be
\hat{\kappa}^b= A_{\rm prediction} \kappa + \cdots \ ,
\ee
where $A_{\rm prediction}=\sum_{i} w^b_{i} g_i^{\rm true}\approx \sum_{i} w^b_{i} g_i$.
On the other hand from the cross correlation analysis for $\hat{\kappa}^b$, 
we can get the constraints on $A$.
The deviation of $A$ from 1 indicates the bias caused by the errors in $g$.
We realize the Gaussian noise for $g_i$ of each flux bin independently, and at three noise levels $5\%, 10\%, 15\%$.
The results are shown in the bottom rows in Table \ref{tab:robust}.
For the noise level of $10\%$ and $15\%$, $A$ deviates from 1 significantly.
The bias is consistent with the prediction $A_{\rm prediction}$,  indicating that the approximation $g_i^{\rm true}\approx g_i$ is valid.
This is also applicable to the baseline and internal tests in Sec. \ref{sec:robustest},
where we used $g_i$ to approximate the magnification coefficient and no Gaussian noise was added.
In this case,  $A_{\rm prediction} = 1$, agrees with the constraints on $A$ (see the first 9 rows in Table \ref{tab:robust}).
Although the bias induced by the error in $g$ is significant, the weak lensing signal is still detected at $S/N > 10$ and the intrinsic clustering is well suppressed.
\\
 \indent
To further demonstrate the robustness of the analysis against the errors in $g$,
we make 50 realizations of the Gaussian error in $g$ at the $15\%$ level.
For each realization, we estimate $A_{\rm prediction}$ and get constraint on $A$ from the cross-correlation analysis. 
The comparison of $A$ and $A_{\rm prediction}$ for the 50 realizations is shown in \reffig{fig:A_Apredict_z9-12}.
For different realizations, the errors in $g$ induces different biases in the reconstructed $\hat{\kappa}^b$,
which all agree well with the prediction $A_{\rm prediction}$.
\\
 \indent
 The above analysis demonstrates how the errors in $g$ can bias the reconstructed lensing convergence.
Although constructed with biased $g$, the weak lensing signal can still be detected at $S/N > 10$ and the intrinsic clustering is well suppressed (the bottom rows in Table \ref{tab:robust}).
The bias mainly affects the overall amplitude of the reconstructed lensing convergence, which is reflected in the deviation of $A$ from 1.
Under the assumption that $g_i^{\rm true}$ can be approximated by the logarithmic slope of the luminosity function, the bias in $A$ can be predicted by $A_{\rm prediction}= \sum_{i} w^b_{i} g_i$.
However the assumption will not hold and $g_i^{\rm true}\ne g_i$ if the galaxy survey selection is more complex than the flux-limited case.
It is discussed in \cite{Wietersheim-Kramsta_Joachimi_van} and \cite{JElvinPoole2022DarkES} that galaxy selections, such as color, position and shape, can lead to biases in the magnification coefficient.
This would be the main source of the bias in $A$ for the reconstructed lensing convergence for $ 0.5 < z_\kappa < 0.8$ if no other systematics are present, 
which we present in section \ref{sec:0.5_0.8}.
}

\begin{figure}
\centering
\includegraphics[width=\columnwidth]{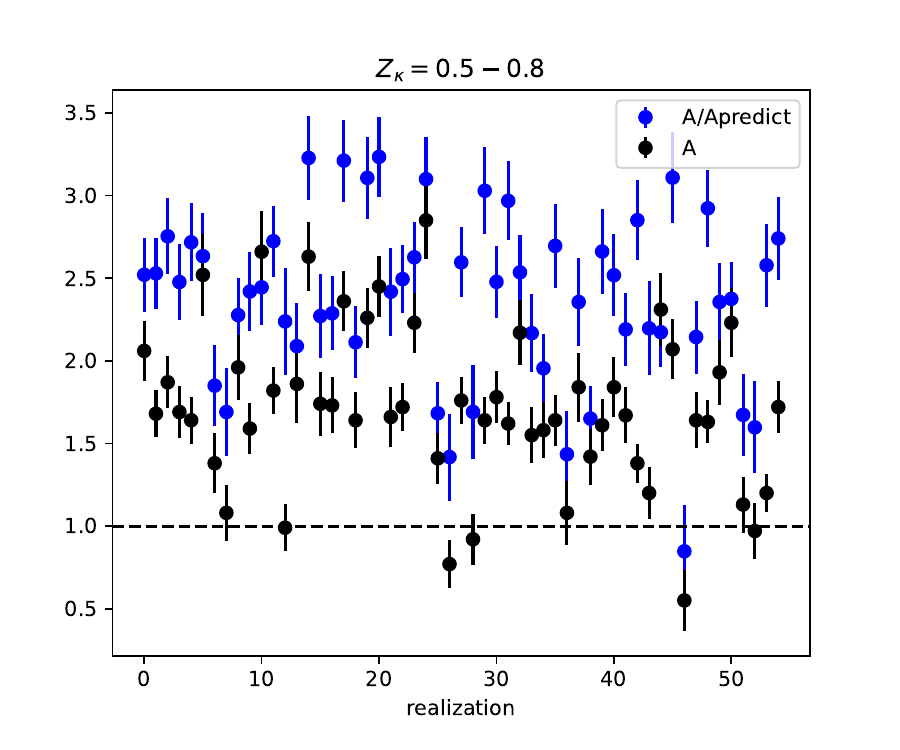}
\caption{ Same as \reffig{fig:A_Apredict_z9-12}, but for the photo-$z$ range $0.5 < z_\kappa < 0.8$.
In this case the bias in $A$ differs from the prediction $A_{\rm prediction}$, 
which implies $g_i$, defined by the logarithmic slope of the luminosity function,
is not a good approximation for the true magnification coefficient for $ 0.5 < z_\kappa < 0.8$.
}
\label{fig:A_Apredict_z5-8}
\end{figure}


\subsection{Reconstruct the intrinsic clustering while suppressing the lensing signal}\label{sec:clustering_only}
{In this part, we aim to show that the cross correlation signal parameterized by $A$ are from the lensing signal, i.e, 
the convergence-shear cross correlation, not from the intrinsic clustering-shear cross correlation.
We have used the condition $\sum w_i g_i=1$ to extract the lensing convergence signal in the reconstruction precedure and 
the condition $\sum w_i =0$ to suppress the intrinsic clustering.
We can also do conversely, i.e., 
we can use the condition $\sum w_i =1$ to extract the intrinsic clustering and the condition $\sum w_i g_i=0$ to suppress the lensing signal.
We minimize the shot noise and get the reconstruction.
In this case, the reconstructed fields are expected to be dominated by the intrinsic clustering and the lensing signal is suppressed,
which can be characterized by the parameter $A$ and $\epsilon$.
We perform the cross-correlation analysis for the reconstructed field to get the constraints on $A$ and $\epsilon$. 
The results are shown in the last row in Table \ref{tab:robust}.
As expected, we get $A\sim 0$ and $\epsilon >0$,
controverting the results of the baseline set ($\sum w_i g_i=1,\sum w_i =0$).
For this construction, the signal is the intrinsic clustering and the residual noise is the lensing convergence.
The residual is tiny, i.e., $A/\sigma_A<1$, which indicates the $g_i$ estimated by the logarithmic slope of the luminosity function is reliable for this case.
}

\subsection{Results for lensing convergence reconstructed at $0.5 < z_\kappa < 0.8$}\label{sec:0.5_0.8}
{The above analysis is for the lensing convergence reconstructed at $0.9 < z_\kappa < 1.2$, 
which presents robustest against various tests.  
In this section, we summarize the main results for the lensing convergence reconstructed at $0.5 < z_\kappa < 0.8$.
The magnitude cuts for the three bands are $m_g=22.5, m_r=22.0, m_z=21.2$.
They are determined by the same strategy as $0.9 < z_\kappa < 1.2$, i.e., $\sim 0.5$ mag fainter than peak of the luminosity function.
The average galaxy surface number density is $n_{\rm gal}=2.0$ arcmin$^{-2}$.
Besides, all the other parameters are kept same as $0.9 < z_\kappa < 1.2$.
The summary of the results is shown in Table \ref{tab:robust5-8}, including the baseline set and all the tests.
The lensing signal is detected at $S/N > 10$ with the intrinsic clustering well suppressed. 
And it shows the robustness against different choices of $f^{\rm threshold}, N_{\rm F}$, 
and magnitude cuts and presents increased S/N with the imaging systematics mitigation.
However there are three main differences from the results of $0.9 < z_\kappa < 1.2$.
First, the lensing amplitude $A$ of the baseline has bias of a factor of $\sim 2$.
Second, regarding the impact of introducing Gaussian noise in $g$ on the constraints of $A$,
the biases in $A$ differ from the predicted values ($A_{\rm prediction}$).
\reffig{fig:A_Apredict_z5-8} presents the comparison between $A$ and $A_{\rm prediction}$ for 50 realizations of Gaussian noise in $g$ at the $15\%$ level,
all of which indicate the invalidity of $A_{\rm prediction}$.
Third, the intrinsic clustering reconstruction ($\sum w_i = 1$) while suppressing the lensing magnification ($\sum w_i g_i=0$) returns a significant residual in $A$ ($A/\sigma_A\sim -3$).
As a comparison, the residual is tiny ($A/\sigma_A\sim 0$) for the lensing convergence reconstructed at $0.9 < z_\kappa < 1.2$.
The three anomalies implies that the logarithmic slope of the luminosity function may not be a good approximation for the magnification coefficient for $ 0.5 < z_\kappa < 0.8$.
How the galaxy survey selection affects the magnification coefficient is discussed in \cite{Wietersheim-Kramsta_Joachimi_van} and \cite{JElvinPoole2022DarkES}.
The true magnification coefficient can be different from $g_i$ defined by the logarithmic slope of the luminosity function by a factor of up to $\sim 2-3$ \citep{Wietersheim-Kramsta_Joachimi_van,JElvinPoole2022DarkES}.
Another potential factor is the impact of intrinsic clustering. However, we have incorporated the term $\epsilon \xi_{\rm m\gamma}$ into our model (see Eq.\ref{eq:xikg_decom}) to address this.
Additionally, inaccuracies in the photo-$z$ model could introduce systematics into the theoretical templates. We tested various assumed values of $\sigma_z$ and found that these variations affect the results insignificantly (see Table \ref{tab:robust_phz}).
}
\begin{table*}
\begin{center}
\begin{tabular}{lllllll}
\hline\hline
{} &                           $A$ &                      $\epsilon$ & $\chi^2_{\rm min}$ & $\sqrt{\chi^2_{\rm null}-\chi^2_{\rm min}}$ & $A/\sigma_A$ \\
\midrule
\hline
baseline $\sigma_z=0.5$ &  2.55$\pm$0.25(2.59$\pm$0.28) &  -0.00$\pm$0.02(-0.01$\pm$0.02) &         26.6(29.4) &                                  17.5(15.8) &    10.2(9.2) \\
$\sigma_z=0.25$         &  2.51$\pm$0.22(2.63$\pm$0.25) &  -0.01$\pm$0.02(-0.02$\pm$0.02) &         24.9(32.9) &                                  16.8(15.8) &   11.4(10.5) \\
$\sigma_z=0.75$         &  2.60$\pm$0.31(2.74$\pm$0.35) &  -0.00$\pm$0.03(-0.02$\pm$0.03) &         26.5(35.4) &                                  17.8(14.8) &     8.4(7.8) \\
$\sigma_z=0.1$          &  2.62$\pm$0.39(3.03$\pm$0.48) &  -0.00$\pm$0.03(-0.04$\pm$0.04) &         28.5(37.0) &                                  17.3(14.7) &     6.7(6.3) \\
\hline
\end{tabular}
\end{center}
    \caption{ Impact of $\sigma_z$ in the photo-$z$ modeling on the cross-correlation analysis for the lensing convergence reconstructed at $0.5 < z_\kappa < 0.8$.
    The baseline set has $\sigma_z=0.5$.
    For the test sets, the only difference is the assumed value of $\sigma_z$.
    The constraints on $A$ and $\epsilon$ are consistent for different assumed values of $\sigma_z$.
      }    \label{tab:robust_phz}
\end{table*}

\section{SUMMARY and discussion}\label{sec:summary5}
Based on cosmic magnification, we reconstruct weak lensing convergence map from the DECaLS DR9 galaxy catalog. 
Notably, this represents the first instance of a directly measured lensing map from magnification, covering a quarter of the sky.
It is done by weighing overdensity maps of galaxies in different magnitude bins and different photometry bands. 
To test validity of the reconstruction, we make cross correlation to the galaxy shear measurement and compare to the theoretical model. 
We find that the galaxy intrinsic clustering is well eliminated by our method and we get $\sim 22\sigma$ detection of the convergence-shear cross correlation.

{In the APPENDIX \ref{sec:gal_x_shear} we directly measure the cross-correlation between the background galaxies and the foreground shears.
The results indicate the true magnification coefficient $g$ can deviate from the flux-based estimation, which assumes the galaxy catalog is a flux-limited sample.
This is more robustly investigated by \cite{Wietersheim-Kramsta_Joachimi_van} and \cite{JElvinPoole2022DarkES}, who use simulations to model the true survey selection functions.
\\
\indent
\red{Errors in the magnification coefficient $g$ can bias the reconstructed lensing convergence.
In Sec. \ref{sec:robust_g}, we introduce Gaussian noise in $g$ and find that it biases the overall amplitude of the reconstructed lensing convergence.}
\\
\indent
The overall amplitude bias observed in the reconstructed lensing convergence at $0.9 < z_\kappa < 1.2$ is small, $\sim 0.9$.
But it is significant for the reconstructed lensing convergence at $0.5 < z_\kappa < 0.8$, $\sim 2$.
To validate and address the bias observed in the reconstructed lensing convergence that originates from systematics in $g$, we need to perform forward modeling of the DR9 galaxy selection function. 
This is our next step in the research project.
}

Our primary focus is on the cross-correlation analysis since the auto-correlation of the reconstructed convergence map is significantly affected by the weighted shot noise \citep{2023arXiv230615177M}. 
Existing data such as DES \citep{DES2016} and HSC \citep[][]{HSC2018} have greater survey depth and higher galaxy number than DECaLS. 
We plan to analyze these data to improve the magnification based lensing reconstruction.
It will enable us to thoroughly account for the model's cosmology dependence, more realistic modeling of the photo-$z$ distribution, the intrinsic alignment of galaxies, and potential contamination from the extragalactic dust.




%
     
\section*{Acknowledgements}
The authors thank Jun Zhang for useful discussions.
This work is supported by National Science Foundation of China (11621303, 12273020), the National Key R\&D Program of China (2020YFC2201602), the China Manned Space Project (\#CMS-CSST-2021-A02 \& CMS-CSST-2021- A03), and the Fundamental Research Funds for the Central Universities.

\bibliography{mybib}

\section{APPENDIX}

\subsection{Weiner filter}
For the spherical harmonic coefficients $\hat\kappa_{\ell m}$ estimated by Healpix,  the Weiner filter is calculated by
\begin{equation}
    \label{eq:wiener}
    \hat{\kappa}^{\rm WF}_{\ell m} = \frac{C_\ell^{\rm signal}}{C_\ell^{\rm signal}+C_\ell^{\rm noise}} \hat\kappa_{\ell m} \ .
\end{equation}
Where $C_\ell^{\rm signal}$ is the auto power spectrum of the lensing signal and 
$C_\ell^{\rm noise}$ is that of the noise.
We calculate $C_\ell^{\rm signal}$ from the Planck 2018 cosmology combined with the bestfit amplitude $A^{\rm bestfit}$.  
To take into account the impact of survey geometry, we generate 100 full-sky maps of the power spectrum $(A^{\rm bestfit})^2 C^{\kappa\kappa}_{\ell,\rm Planck} $. 
We then apply the mask to these maps and obtain 100 power spectra with the DECaLS mask.
We use the average one to get $C_\ell^{\rm signal}$.

\begin{figure*}
    \centering
    \includegraphics[width=\textwidth]{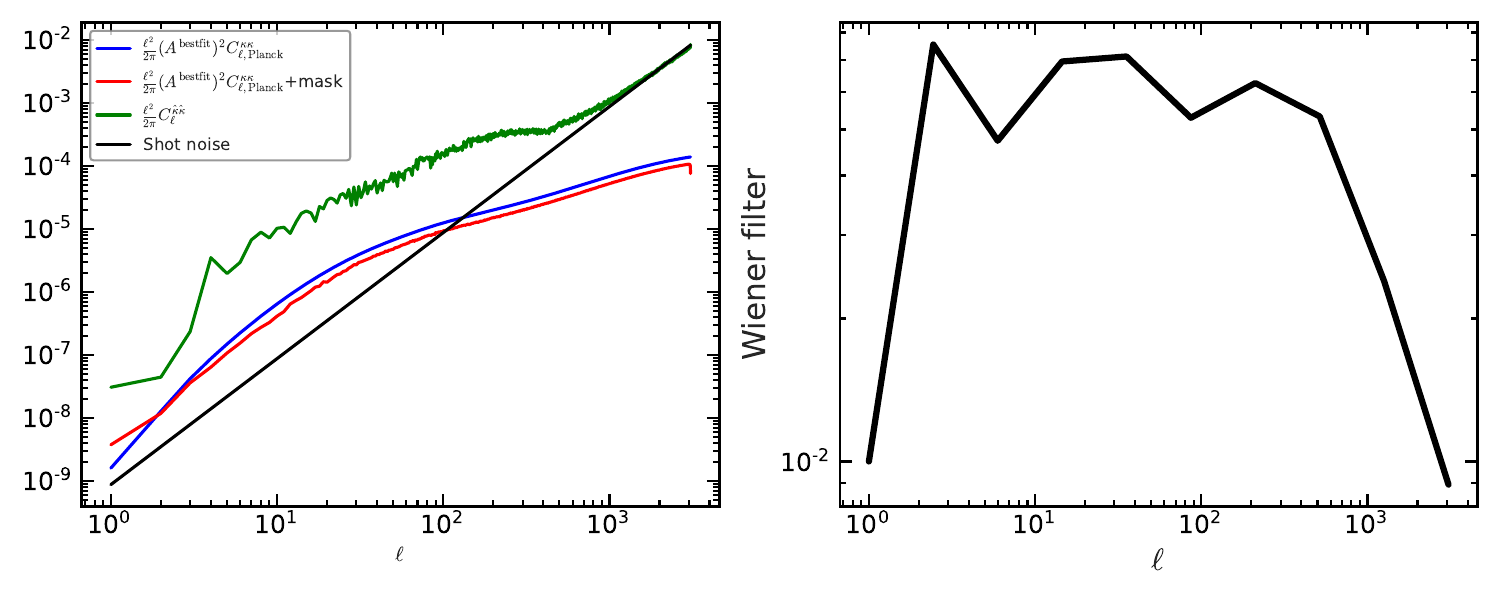}
    \caption{Left: The angular power spectrum of the weak lensing map.
    The blue line is calculated from the Planck 2018 cosmology combined with the bestfit lensing amplitude $A^{\rm bestfit}$,
    while after applying the DECaLS mask the calculation is shown with the red line.
    The green line is the measurement from the reconstructed convergence map. 
    The black line represents the shot noise.
    Right: The associated Weiner filter calculated from the angular power spectrum, by Eq.\eqref{eq:wiener}.
    The lensing convergence map after applying the Weiner filter is shown in \reffig{fig:kappamap}.
    }
    \label{fig:wiener}
\end{figure*}

We calculate $C_\ell^{\rm signal}+C_\ell^{\rm noise}$ by the auto power spectrum of the reconstructed lensing map $\hat\kappa$, which is
\begin{equation}
    C_\ell^{\hat\kappa\hat\kappa} = \frac{1}{2\ell+1}\sum\limits_{m=-\ell}^\ell \hat\kappa_{\ell m}^* \hat\kappa_{\ell m} \ .
\end{equation}
Take the reconstructed lensing map of the baseline set (the first row in Table \ref{tab:robust}) as an example, the results of the power spectrum is shown in \reffig{fig:wiener}.
The auto power spectrum of the reconstructed lensing map is overwhelmed by the shot noise at $\ell>50$,
which agrees with the results in \cite{2023arXiv230615177M}.
This is the reason why we have used the cross correlation to increase the S/N.
The associated Weiner filter calculated by Eq.\eqref{eq:wiener} is shown in the right panel of \reffig{fig:wiener} and the lensing convergence map after applying the Weiner filter is shown in \reffig{fig:kappamap}.

\begin{figure*} 
\subfigure{\includegraphics[width=2\columnwidth]{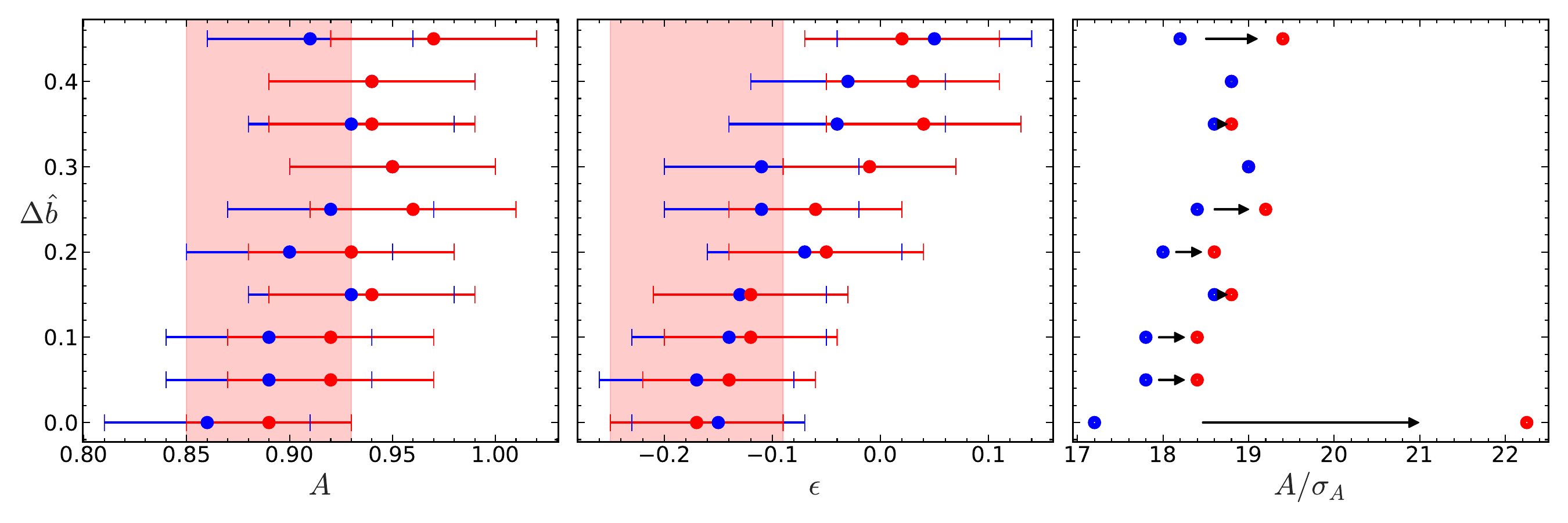}}
\subfigure{\includegraphics[width=2\columnwidth]{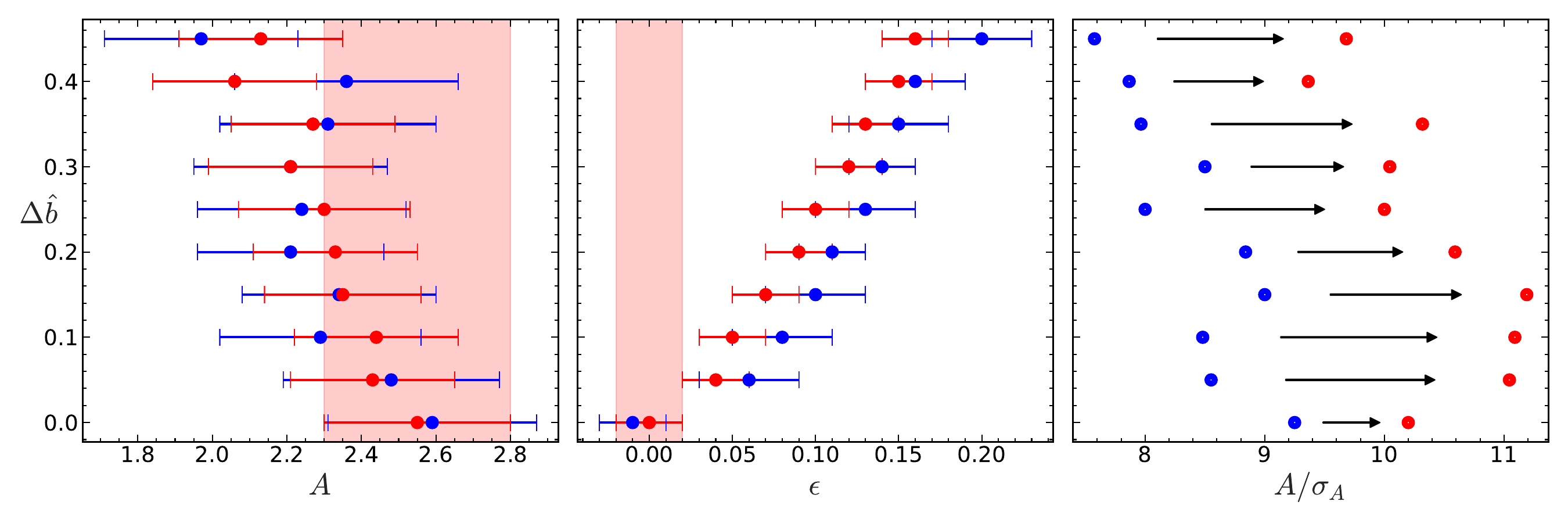}}
\caption{Impact of varying the condition $\sum w_i \hat{b}_i$ on the constraints of $A$ (the left panel), $\epsilon$ (the middle panel) and $A/\sigma_A$ (the right panel). 
The red/blue data points denotes the results with/without imaging systematics mitigation. 
The top/bottom panels are for the lensing convergence reconstructed at $0.9 < z_\kappa < 1.2$ / $0.5 < z_\kappa < 0.8$.
In the left/middle panels, the shaded regions represent the $1 \sigma$ error in the constraint on $A$ / $\epsilon$ for the baseline set,
which corresponds to $\Delta\hat{b}=0$ and includes imaging systematics mitigation.
The constraints on $A$ are stable for different $\Delta\hat{b}$ for all the cases.
The constraints on $\epsilon$ exhibit dependence on $\Delta\hat{b}$, and tend to have a linear relation with $\Delta\hat{b}$.
The right panels compare the S/N $\equiv A/\sigma_A$ before and after the imaging systematics mitigation, for different $\Delta\hat{b}$.
The S/N show increase after the mitigation for all the investigated cases of $\Delta\hat{b}$.
\label{fig:Ae_varyb}}
\end{figure*}

\subsection{Impact of the galaxy intrinsic clustering on the lensing reconstruction}
To eliminate the average galaxy clustering in the lensing reconstruction, we initially employed the condition $\sum w_i = 0$. However, we can reevaluate the impact of average galaxy clustering by modifying the condition to
\be \label{eqn:wb_beyond} \sum_{i} w_{i}\hat{b}_i=0\ . \ee 
Here, the values of $\hat{b}_i$ are artificially selected.
Setting all $\hat{b}_i$'s equal to a constant reduces to the original condition.
Selecting forms that $\hat{b}_i$'s deviate from a constant will enhance the impact of the galaxy clustering. 
The results of $\epsilon$ are expected to be dependent on the form of $\hat{b}_i$'s, while those of $A$ are not.
We use a simple form of $\hat{b}_i$ as a demonstration.
For each band, we set $\hat{b}_i$ for the four flux bins to be
\be\label{eq:Deltap}
1-2\Delta \hat{b},1-\Delta \hat{b},1+\Delta \hat{b},1+2\Delta \hat{b}
\ee
The only parameter is $\Delta \hat{b}$, which we set to be identical for each band.
We sample $\Delta \hat{b}$ in the range $[0,0.45]$ with bin width $0.05$.
For each $\Delta \hat{b}$ we repeat the lensing reconstruction procedure and conduct the cross-correlation analysis to obtain the constraints on $A$ and $\epsilon$.

\reffig{fig:Ae_varyb} shows the impact of $\Delta \hat{b}$ on the results of constraints.
The best fit of $A$ is insensitive to the choice of $\Delta \hat{b}$, as expected, showing the stability of the reconstruction.
However, the results of $\epsilon$ exhibit significant dependence on $\Delta \hat{b}$, and tend to have a linear relation with $\Delta \hat{b}$.
Its deviation from the baseline is significant when $|\Delta \hat{b}|$ differs considerably from zero. 
The tendency is more pronounced for $0.5 < z_\kappa < 0.8$, because of the higher constraining power on $\epsilon$ compared to $0.9 < z_\kappa < 1.2$.
Therefore by varying the condition to Eq. \eqref{eqn:wb_beyond}, we revealed the evidence of galaxy clustering which contaminates the cosmic magnification.
\reffig{fig:Ae_varyb} also shows that the imaging mitigation enhances S/N for all the investigated cases of $\Delta \hat{b}$.

\subsection{Cross-correlation between the foreground shear and the background galaxies}
\label{sec:gal_x_shear}
The lensed galaxies at high redshift are magnified by foreground structures, detectable through the cross-correlation of background galaxies and foreground shear. For background galaxies, we use subsamples similar to those in the lensing reconstruction, defined within $0.9 < z_{g,{\rm photo}} < 1.2$. The definitions are summarized in Table \ref{tab:sub-sample}.

For foreground shear, we use the shear measurements described in Sec. \ref{sec:data3}, considering two photo-$z$ bins: $0.4 < z_{\gamma,{\rm photo}} < 0.45$ and $0.3 < z_{\gamma,{\rm photo}} < 0.4$. We measure the cross-correlation between shear and galaxy subsamples. The estimator for this cross-correlation is

$$
\xi^{\mathrm{\gamma g}}(\theta)=\frac{\sum_{\mathrm{ED}} \gamma_{\mathrm{E}}^{+} \mathbf{w}_{\mathrm{D}}}{\sum_{\mathrm{ER}}\left(1+m_{\mathrm{E}}\right) \mathbf{w}_{\mathrm{R}}}-\frac{\sum_{\mathrm{ER}}  \gamma_{\mathrm{E}}^{+} \mathbf{w}_{\mathrm{R}}}{\sum_{\mathrm{ER}}\left(1+m_{\mathrm{E}}\right) \mathbf{w}_{\mathrm{R}}}\ ,
$$
where \( m_{\mathrm{E}} \) and \( \gamma_{\mathrm{E}}^{+} \) denote the multiplicative bias correction and the tangential shear of the foreground galaxies, respectively. \( \mathbf{w}_{\mathrm{D}} \) and \( \mathbf{w}_{\mathrm{R}} \) are the imaging systematic weights. The \(\Sigma\)-summations are calculated for all ellipticity-density (ED) and ellipticity-random (ER) pairs.

\begin{figure*}
    \centering
    \includegraphics[width=\textwidth]{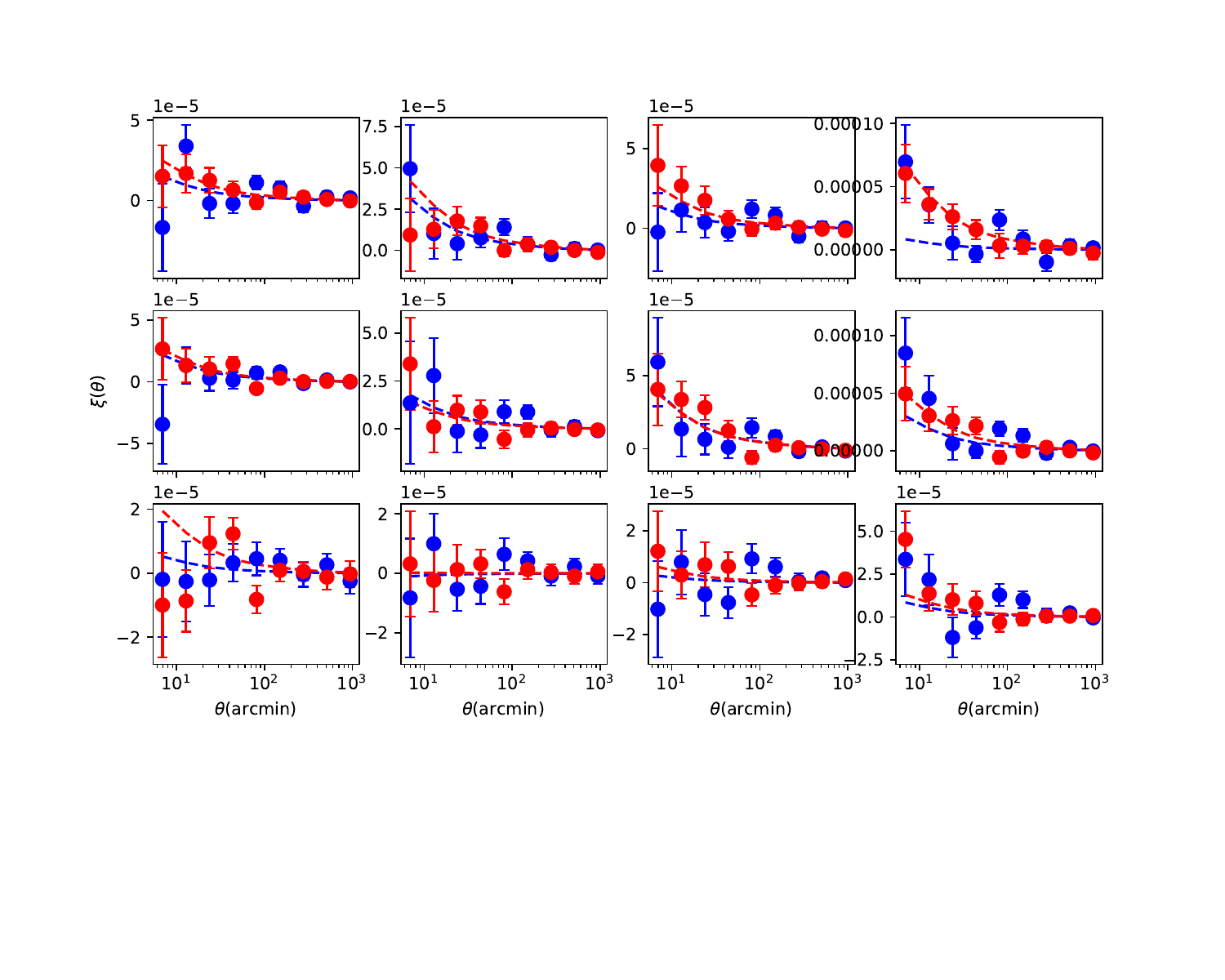}
    \caption{Measured cross-correlation between the background galaxies and the foreground shears.
    The background galaxies are the DECaLS galaxy subsamples at $0.9 < z_{g,{\rm photo}} < 1.2$, whose definition are summarized in Table \ref{tab:sub-sample}. 
    Each panel shows the results for a different subsample.
    For the foreground shears, two shear photo-$z$ bins are considered, $0.4 < z_{\gamma,{\rm photo}} < 0.45$ and $0.3 < z_{\gamma,{\rm photo}} < 0.4$,
    which are represented by the blue and red data points, respectively.
    The dashed lines are the fitting to the measurements, 
    which we only consider the lensing signal term (i.e. the convergence-shear cross-correlation) and ignore the possible contamination from the intrinsic clustering.
    }
    \label{fig:gglensing1}
\end{figure*}

\begin{figure*} 
    \subfigure{\includegraphics[width=1\columnwidth]{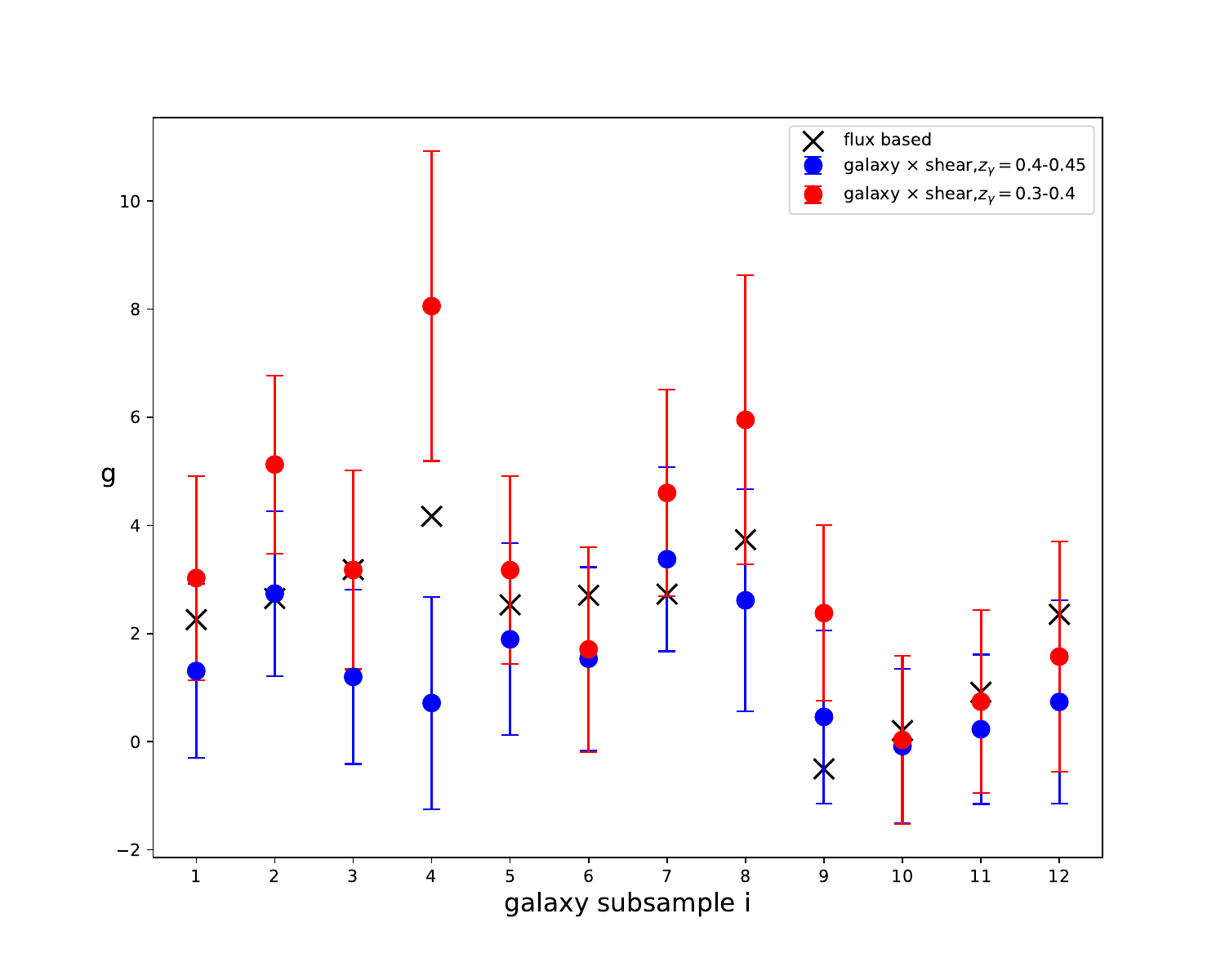}}
    \subfigure{\includegraphics[width=1\columnwidth]{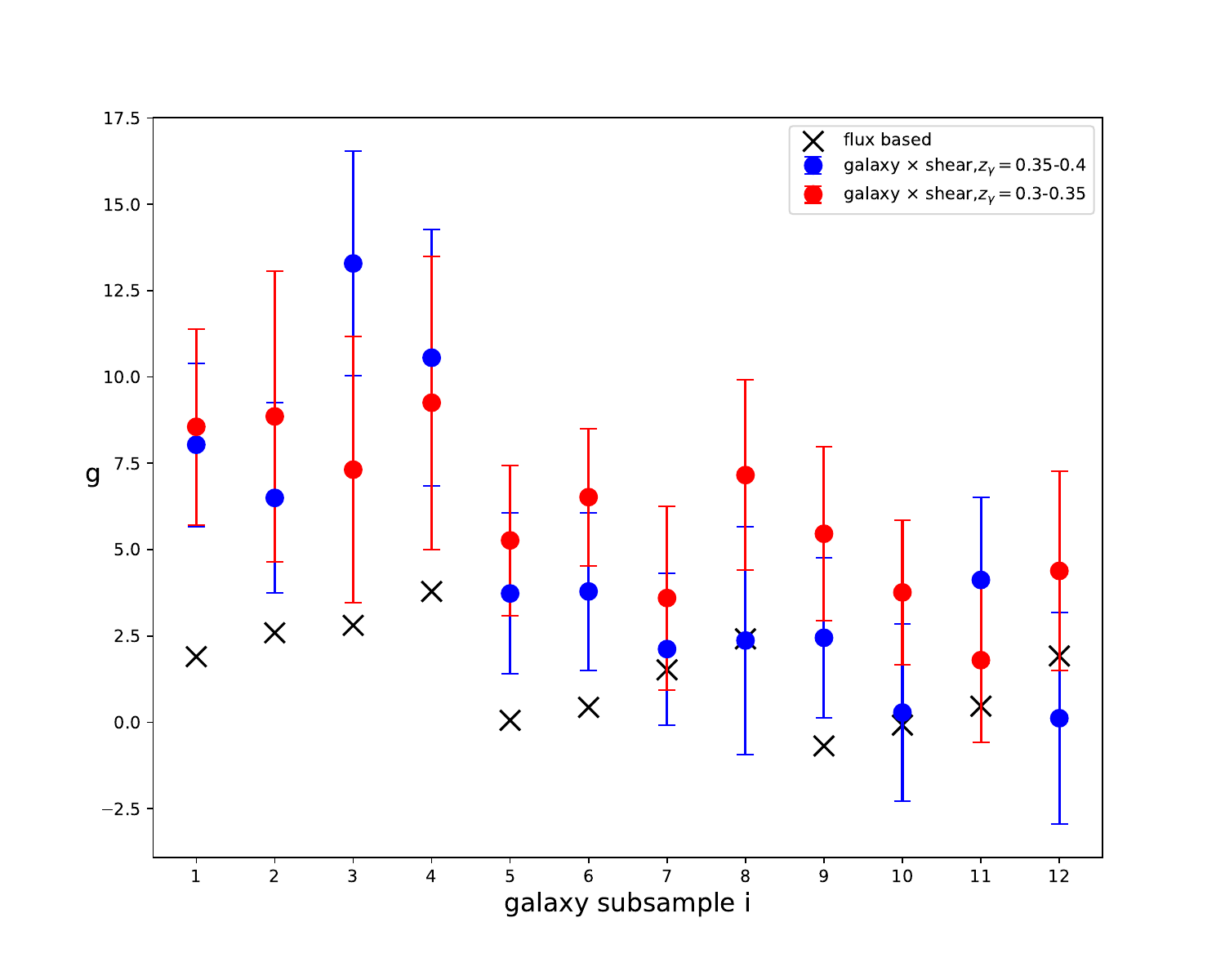}}
    \caption{The the results of $g_i$ for the DECaLS galaxy subsamples at $0.9 < z_{g,{\rm photo}} < 1.2$ (the left panel) and $0.5 < z_{g,{\rm photo}} < 0.8$ (the right panel).
    The blue and red data points denote the results for the lens galaxies at different photo-$z$ bins.
    The cross data points are measured from the flux-based method, i.e., the logarithmic slope of the luminosity function.
    }
    \label{fig:ggfit}
    \end{figure*}

The measurement results are shown in \reffig{fig:gglensing1}. By fitting these measurements, we estimate the magnification coefficient \( g_i \) for each subsample. The model for the cross-correlation is

$$
\xi_{\rm th}^{\gamma g} = g_i \langle\kappa\gamma\rangle\ , 
$$
where the model of  $\langle\kappa\gamma\rangle$ follows \eqref{eq:kg:hankel} and is calculated under the Planck 2018 cosmology.
Given that the photo-$z$ of the foreground are much lower than those of the background, we only considered the convergence-shear cross correlation in the modeling, ignoring possible contamination from intrinsic clustering. 
The fitted \( g_i \) results are shown in \reffig{fig:ggfit}, where we compared these results to those obtained using the flux-based method, which involves the logarithmic slope of the luminosity function. For the galaxy subsamples, the \( g_i \) results are consistent with the flux-based method, validating the approximation of the magnification coefficient by the logarithmic slope of the luminosity function in this case.

\begin{figure*}
    \centering
    \includegraphics[width=\textwidth]{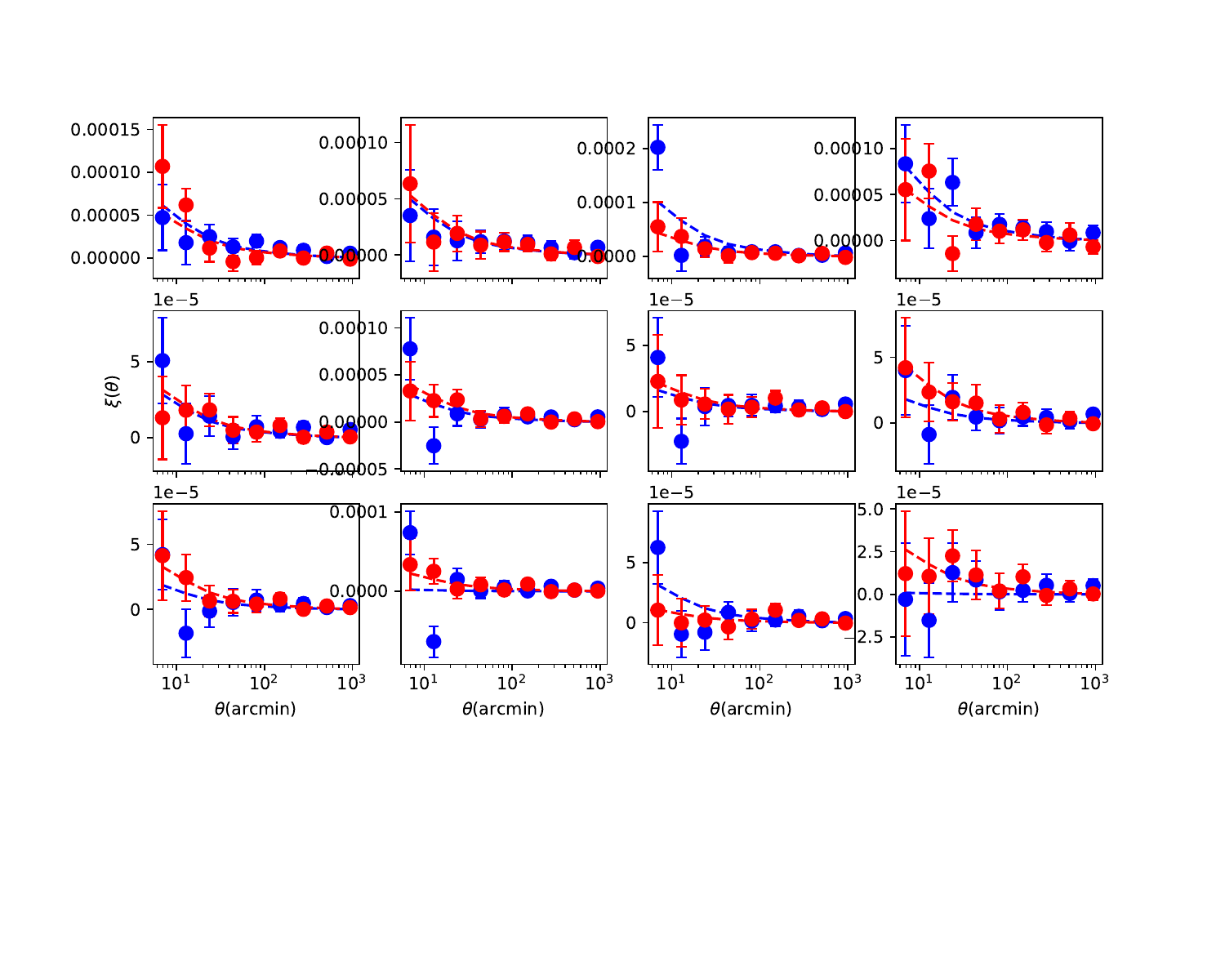}
    \caption{Same as \reffig{fig:gglensing1}, but for background galaxies at $0.5 < z_{g,{\rm photo}} < 0.8$ and foreground shears at $0.35 < z_{\gamma,{\rm photo}} < 0.45$ and $0.3 < z_{\gamma,{\rm photo}} < 0.35$.
    }
    \label{fig:gglensing2}
\end{figure*}

We conduct the same analysis for the background/galaxy subsamples at $0.5 < z_{g,{\rm photo}} < 0.8$ and the foreground/shear at $0.35 < z_{\gamma,{\rm photo}} < 0.45$ and $0.3 < z_{\gamma,{\rm photo}} < 0.35$.
 The results of the cross-correlation measurements are presented in \reffig{fig:gglensing2}, 
 and the fitted $g_i$ values are shown in \reffig{fig:ggfit}. 
 The fitted $g_i$ results are systematically lower than those obtained using the flux-based method. 
 This trend aligns with the bias of a factor of approximately 2 in the lensing amplitude $A$ for the reconstructed lensing convergence at $0.5 < z_\kappa < 0.8$ (see Table \ref{tab:robust5-8}).
  However, this discrepancy might also be attributed to contamination from intrinsic clustering. 
  This is suggested by the discrepancy in the fitted $g_i$ results for the two shear photo-$z$ bins (the red and blue data points). 
 However, the low S/N of the cross-correlation measurements prevent us from fully considering the impact from the intrinsic clustering. 
To further investigate this issue, we need improved survey data. This could include data from DES and HSC, which offer higher galaxy number densities, or spectroscopic data from DESI, which would provide more definitive redshift space separation between foreground and background galaxies.
 This constitutes the next step in our research project.

\label{lastpage}
\end{document}